\documentclass[
aip,
jcp,
floatfix,
 amsmath,amssymb,
preprint,%
 reprint,%
]{revtex4-1}

\usepackage{graphicx}
\usepackage{dcolumn}
\usepackage{bm}

\usepackage[utf8]{inputenc}
\usepackage[T1]{fontenc}
\usepackage{mathptmx}
\usepackage{etoolbox}
\usepackage{xcolor}
\usepackage{siunitx}

\newcommand{\beq}{\begin{equation}}
\newcommand{\eeq}{\end{equation}}

\newcommand{{\etal}} {\textit{et al.}}
\newcommand{\mrm}{\mathrm}
\newcommand{\rL}{\rho_\mrm{L}}
\newcommand{\muL}{\mu_\mrm{L}}

\newcommand{\rV}{\rho_\mrm{V}}

\newcommand{\rav}{\rho_0}
\newcommand{\dav}{\delta_0}
\newcommand{\ds}{\delta_\mathrm{0,sp}}
\newcommand{\de}{\delta_\mathrm{0,eq}}
\newcommand{\done}{\delta_\mathrm{0}^\ast}
\newcommand{\donerel}{\delta_\mathrm{rel}^\ast}
\newcommand{\xs}{x_\mathrm{sp}}
\newcommand{\xe}{x_\mathrm{eq}}
\newcommand{\xc}{x_\mathrm{c}}
\newcommand{\xbub}{x_\mathrm{bub}}
\newcommand{\ve}{v_\mathrm{e}}
\newcommand{\vc}{v_\mathrm{c}}
\newcommand{\vbub}{v_\mathrm{bub}}

\newcommand{\Th}{T_\mathrm{h}}

\newcommand{\kB}{k_\mathrm{B}}
\newcommand{\kb}{k_\mathrm{B}}
\newcommand{\Eb}{E_\mathrm{b}}
\newcommand{\phic}{\phi_\mathrm{c}}
\newcommand{\Zliq}{Z_\mathrm{liq}}
\newcommand{\pliq}{p_\mathrm{liq}}
\newcommand{\kliq}{k_\mathrm{liq}}
\newcommand{\tauliq}{\tau_\mathrm{liq}}
\newcommand{\Zbub}{Z_\mathrm{bub}}
\newcommand{\pbub}{p_\mathrm{bub}}
\newcommand{\kbub}{k_\mathrm{bub}}
\newcommand{\taubub}{\tau_\mathrm{bub}}

\makeatletter
\def\@email#1#2{%
 \endgroup
 \patchcmd{\titleblock@produce}
  {\frontmatter@RRAPformat}
  {\frontmatter@RRAPformat{\produce@RRAP{*#1\href{mailto:#2}{#2}}}\frontmatter@RRAPformat}
  {}{}
}%
\makeatother

\begin{document}

\preprint{AIP/123-QED}

\title{Phase fluctuations in a confined fluid}
\author{Frédéric Caupin}
\email{frederic.caupin@univ-lyon1.fr}
 \affiliation{Institut Lumi\`ere Mati\`ere, Universit\'e Claude Bernard Lyon 1, CNRS, Institut Universitaire de France, F-69622, Villeurbanne, France}

\author{Alberto Zaragoza}%
\affiliation{Departamento de Matem\'aticas y Ciencias de Datos, Universidad San Pablo-CEU, CEU Universities, Madrid, Spain
}%

\author{Miguel A. Gonzalez}
\affiliation{Chemical, Energy and Mechanical Technology Department, ESCET. Universidad Rey Juan Carlos, c/ Tulipán s/n, 28933 Móstoles, Madrid, Spain.}

 \author{Chantal Valeriani}
 \affiliation{Departamento de Estructura de la Materia, F\'sica T\'ermica y Electr\'onica, Universidad Complutense de Madrid, Madrid, Spain}

\date{\today}

\begin{abstract}
Fluid phase equilibrium depends on the external constraints imposed on a system. In a closed system with fixed volume, depending on the average density, a vapor bubble may be stable, metastable, or unstable, with respect to the homogeneous liquid phase. In the case where the bubble is metastable, we study its lifetime, i.e. the average waiting time needed to observe bubble collapse, and the corresponding lifetime of the homogeneous liquid. For the smallest systems, we predict the possibility to observe phase flipping, when the fluid oscillates between states with and without bubble. We provide an example of phase flipping in a simulation of a Lennard-Jones fluid.
\end{abstract}

\maketitle

\section{\label{sec:Introduction}Introduction}

Many phenomena involve the crossing of an energy barrier by a particle or a many-body system. Examples include chemical reactions, radioactive alpha decay, protein folding~\cite{Rollins_general_2014,Lyons_quantifying_2024}, microspheres in optical traps~\cite{McCann_thermally_1999,Chupeau_optimizing_2020,Lyons_quantifying_2024}, nucleation in metastable phases~\cite{Debenedetti_metastable_1996,Caupin_nucleation_2026}, etc{\ldots} A particular case of the last is cavitation, i.e. the appearance of a bubble in a liquid stretched to a density below its saturated vapor pressure value. In classical nucleation theory (CNT),~\cite{Gibbs_equilibrium_1878,Volmer_Keimbildung_1926,Fisher_fracture_1948,Debenedetti_metastable_1996}
the minimum work needed to form spherical bubbles results from the competition between the volume energy gained by replacing the metastable liquid by vapor, and the energy cost associated with the interface between two phases. This results in free-energy barrier $\Eb$ to be overcome, which happens at a rate $J\propto \exp [ - \Eb/( \kB T)]$. For water near ambient temperature in the absence of impurities, spontaneous cavitation requires large negative pressures to occur, beyond \qty{100}{MPa}~\cite{Zheng_liquids_1991,Azouzi_coherent_2012}. To go beyond CNT, advanced simulation techniques are employed; for instance, umbrella sampling applied to stretched water gave nucleation rates consistent with experiments~\cite{Menzl_molecular_2016}.

Here we address the effects of confinement on the kinetics of phase change between liquid and vapor. Porous materials are ubiquitous in nature, and may lead to capillary condensation. This results in an \textit{open} confinement, when the condensed liquid can exchange molecules with the outer vapor. We are interested in \textit{closed} confinement, when the fluid is trapped inside a host material preventing external exchange. Such fluid inclusions are found in minerals\cite{Hurai_geofluids_2015}, and widely studied to obtain information about the conditions at which the mineral formed, with useful applications to paleotemperature reconstruction~\cite{Roedder_fluid_1984,Kruger_liquid_2011,Guillerm_physical_2025}. A key parameter in these studies is the temperature $\Th$ at which a vapor bubble, initially present in the inclusion, disappears upon heating. In the absence of surface tension, the bubble radius would reach zero, and from $\Th$ one would directly obtain the average fluid density in the inclusion. However, the existence of surface tension and the finite compressibility of the liquid result in rich thermodynamic phenomena, such as the possibility for a liquid at negative pressure to remain absolutely stable, thus making the connection between the observed $\Th$ and the actual density uncertain. Previous treatments in the literature were limited to the thermodynamics of the problem, which can only provide bounds on the possible values for $\Th$ at a given fluid density~\cite{Marti_effect_2012,Wilhelmsen_communication_2014,Wilhelmsen_evaluation_2015,Vincent_statics_2017,Caupin_effects_2022,Guillerm_physical_2025}. Here we deal with the kinetics, in order to predict at which conditions the bubble is expected to disappear. We also identify an interesting regime, for which the system exhibits reversible phase fluctuations between states with and without bubble. Although occurring only for cavities too small to be observed under the microscope, this effect can be studied in simulations, for which we provide an example.

This article is organized as follows. In Section~\ref{sec:methods}, we present the theoretical model (Section~\ref{sec:model}) and the simulation details (Section~\ref{sec:simuls}). The results are shown in Section~\ref{sec:results}, and discussed in Section~\ref{sec:discussion}.

\section{Methods\label{sec:methods}}

\subsection{Model\label{sec:model}}

Several models have considered the thermodynamics of the liquid-vapor transition in a system of fixed volume~\cite{MacDowell_nucleation_2006,Marti_effect_2012,Glavatskiy_effect_2013,Wilhelmsen_communication_2014,Wilhelmsen_evaluation_2015,Vincent_statics_2017,Llamas-Jaramillo_freeenergy_2026}. Here we will follow the model of Ref.~\onlinecite{Caupin_effects_2022}, which we summarize in Section~\ref{sec:F}. We also present  in Section~\ref{sec:rate} how, in order to address the kinetics of the problem, we have adapted the approach developed by Menzl~{\etal} to obtain cavitation rates~\cite{Menzl_molecular_2016}.

\subsubsection{Free energy\label{sec:F}}

In Ref.~\onlinecite{Caupin_effects_2022}, we considered a cavity of volume $V$ filled with a fluid at temperature $T$. The cavity is not restricted to be spherical, but we introduce its equivalent radius $R=[3V/(4\pi)]^{1/3}$. The fluid in the cavity can be homogeneous with density $\rav$, or it can be separated in two phases: a vapor bubble of volume $v$ at density $\rV$, and a liquid of volume $V-v$ at density $\rL$. We assume the liquid wets the cavity walls, so that the bubble is spherical.

To calculate $\Delta F(v)$, the free energy change of the bubble state with respect to the fully liquid state, we use a simplified equation of state based on a linear expansion of the liquid chemical potential:
\beq
\mu_L = \muL^\infty + \frac{1}{{\rL^\infty}^2 \kappa} (\rL - \rL^\infty ) \, ,
\label{eq:mu}
\eeq
where $\kappa$ is the liquid isothermal compressibility at saturated vapor pressure, and quantities corresponding to liquid-vapor equilibrium of an infinite system with a flat interface are indicated with the superscript $\infty$. Equation~\ref{eq:mu}, already used in Ref.~\onlinecite{MacDowell_nucleation_2006}, is a good approximation for moderate changes in density. For water at \qty{20}{\degreeCelsius} for instance, Eq.~\ref{eq:mu} predicts a pressure in good agreement with experiments for densities down to at least $0.97 \rL^\infty$~(Ref.~\onlinecite{Caupin_effects_2022}).

\begin{figure}[t]
\includegraphics[width=0.95\columnwidth]{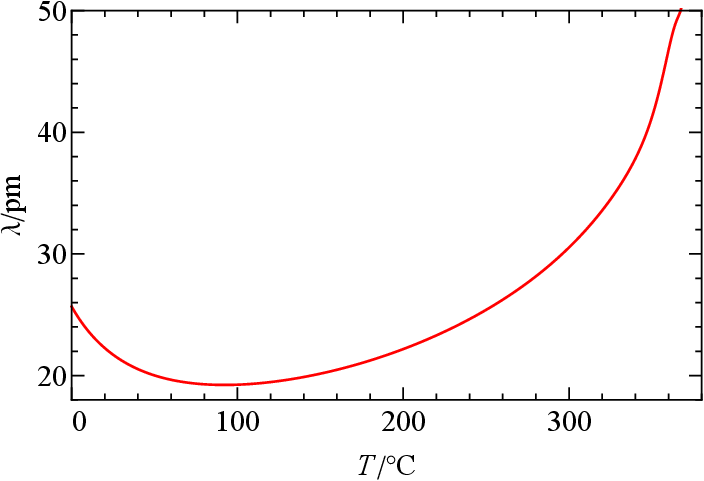}
\caption{Berthelot-Laplace length (Eq.~\ref{eq:BL}) as a function of temperature for pure water.\cite{TheInternationalAssociationforthePropertiesofWaterandSteam_revised_1992,TheInternationalAssociationforthePropertiesofWaterandSteam_revised_2014,TheInternationalAssociationforthePropertiesofWaterandSteam_revised_2018}\label{fig:lambda}}
\end{figure}

Treating the vapor as a perfect gas and assuming $\rV  \ll \rL$ (i.e. $T$ much lower than the critical temperature $T_\mathrm{c}$), we arrived at a simple expression for the non-dimensional free energy change $\phi = 2 \kappa \Delta F/V$:
\beq
\phi = \left( \frac{{\dav}^2}{1-x} -1 \right) x + 9 \epsilon x^{2/3} \, ,
\label{eq:phi}
\eeq
where $\dav=\rav/\rL^\infty$, $x=v/V$, and $\epsilon = \lambda/R$. Here we have introduced the Berthelot-Laplace length:
\beq
\lambda = \frac{2}{3} \gamma \kappa \, ,
\label{eq:BL}
\eeq
where $\gamma$ is the liquid-vapor surface tension. $\lambda$ is a microscopic length: for water it is typically a few tens of \unit{\pm}, see Fig.~\ref{fig:lambda}.

\begin{figure}[t]
\centering
\includegraphics[width=0.95\columnwidth]{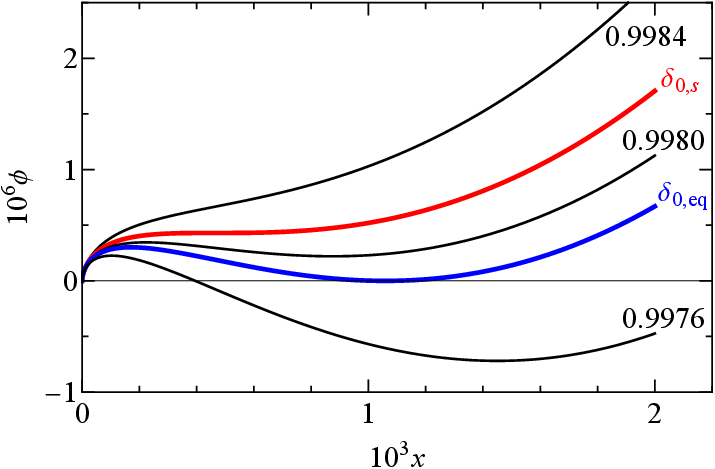}
\caption{Reduced free energy as a function of reduced bubble volume for pure water at $T=\qty{20}{\degreeCelsius}$ (Eq.~\ref{eq:phi}). The black curves show reduced densities $\dav=0.9984$, $0.9980$, and $0.9976$, correspondings to regimes (i), (ii), and (iii), respectively. The red and blue curves correspond to the spinodal ($\ds\approx0.99815$) and binodal ($\de \approx 0.99789$) curves, respectively.
\label{fig:phi}
}
\end{figure}

The value of $\dav$ determines 3 different regimes, as shown in Fig.~\ref{fig:phi}: (i) for $1>\dav >\ds$, $\phi(x)$ is monotonously increasing, and the system can exist only in the fully liquid state; (ii) for $\ds > \dav > \de$, $\phi(x)$ exhibits two minima, corresponding to a stable, fully liquid state, and a metastable state with a bubble; (iii) for $\de > \dav $, $\phi(x)$ exhibits two minima, corresponding to a metastable, fully liquid state, and a stable state with a bubble. The case $\dav=\ds$ is called the bubble \textit{spinodal}, as it corresponds to an inflection point in $\phi(x)$ which renders the bubble state unstable. The case $\dav=\de$ is called the bubble \textit{binodal}, as it corresponds to the conditions at which the free-energy of the homogenous liquid and the stable bubble are equal.

Analytic solutions for $\ds$ and $\de$ and the corresponding reduced bubble volumes $\xs$ and $\xe$ are available, see Ref.~\onlinecite{Caupin_effects_2022} for details. In the limit $\epsilon \ll 1$, the solutions can be expanded to give simple expressions:
\begin{eqnarray}
\label{eq:dsapprox}\ds & = & 1- 4\epsilon^{3/4} + \mathcal{O}(\epsilon^{3/2}) \, ,\\
\label{eq:xsapprox}\xs & = &\epsilon^{3/4} + \mathcal{O}(\epsilon^{3/2}) \, ,
\end{eqnarray}
and
\begin{eqnarray}
\label{eq:deapprox}\de & = & 1- 2 (3 \epsilon)^{3/4} + \mathcal{O}(\epsilon^{3/2}) \, ,\\
\label{eq:xeapprox}\xe & = &(3 \epsilon)^{3/4} + \mathcal{O}(\epsilon^{3/2}) \, .
\end{eqnarray}

In cases (ii) and (iii), $\phi(x)$ reaches a maximum at $\xc$ whose value $\phic$ is involved in the reduced free-energy barrier, that needs to be overcome by thermal fluctuations to allow the system to change state. The reduced volume $\xc$ of the critical bubble is obtained by numerically solving $\mrm{d}\phi/\mrm{d}x=0$, and computing $\phic = \phi(\xc)$. 

For $\dav=\de$, and in the limit $\epsilon \ll 1$, $\xc$ scales as $\xe$, i.e. like $\epsilon^{3/4}$. Keeping only the lowest order in $\mrm{d}\phi/\mrm{d}x$:
\beq
\frac{\mrm{d}\phi}{\mrm{d}x} \approx \frac{2 x^{4/3} - 4 \; 3^{3/4} x^{1/3} \epsilon^{3/4} + 6 \epsilon}{(1-x)^2 x^{1/3}} \,
\label{eq:dphidx}
\eeq
allows obtaining the approximate analytic results:
\beq
\xc \simeq A_x \epsilon^{3/4} \quad \mathrm{and} \quad \phic \simeq A_\phi \epsilon^{3/2} \, .
\label{eq:xcphicapprox}
\eeq
Here
\begin{eqnarray}
\label{eq:Ax}A_x & = & \frac{-4+5 \alpha - \alpha^2}{3^{1/4} \alpha} \, ,\\
\label{eq:Aphi}A_\phi & = & 
\frac{3}{\alpha^2}  \left\{ (2 + \alpha) \beta - 3 \sqrt{3} \alpha^2 \right. \\ \nonumber
& & \left. + 3^{5/6} \alpha^{4/3} \left[ -4+5 \alpha - \alpha^2 \right]^{2/3} \right\} \, .
\end{eqnarray}
with $\alpha = (19 + 3 \sqrt{33})^{1/3}$ and $\beta= 3 \sqrt{3} + \sqrt{11}$, so that $\xc \approx 0.37 \,\epsilon^{3/4} \approx \xe /6.2$ and $\phic \approx 1.4 \,\epsilon^{3/2}$.

\subsubsection{Mean first passage time\label{sec:rate}}

Our goal is to estimate the average time it takes for the system to jump over the free-energy barrier, either from the bubble state to the fully liquid state, or vice-versa.

To this end, we follow a method introduced by Menzl~\textit{et al.}~\cite{Menzl_molecular_2016} to study cavitation. They considered the nucleation of a vapor bubble in an unconfined liquid at temperature $T$, stretched to negative pressure. The fully liquid state is then in a metastable potential well, and a free-energy barrier must be overcome to create a bubble that expands without limit. Kramers' theory~\cite{Kramers_Brownian_1940,Schulten_dynamics_1981} assumes a diffusive motion of the system over the barrier. Using a suitable reaction coordinate $q$ (such as the bubble radius or volume) to describe the free energy $U(q)$, the escape rate $k$ over the barrier is given by :
\beq
k = \frac{\int_\cup \exp \left[-\beta U(q) \right] \,\mrm{d}q}{\int_{\cup} \frac{1}{D(q)}\exp \left[\beta U(q) \right] \,\mrm{d}q} \, ,
\label{eq:k}
\eeq
where $\beta = 1/(\kB T)$, $D(q)$ is the diffusion coefficient, and the symbols $\cup$ and $\cap$ indicate integration over the well and the barrier, respectively. Based on a statistical committor analysis, Menzl~\textit{et al.} found that the volume of the largest bubble $v$ was a good choice for the reaction coordinate $q$. Menzl~\textit{et al.}\cite{Menzl_molecular_2016}  proceeded with calculating the number of cavitation events per unit volume and time, i.e. the nucleation rate:
\beq
J = \frac{\sqrt{-c } D(\vc)}{\sqrt{2\pi \kB T}} \frac{P(\vc)}{V} \, ,
\label{eq:J}
\eeq
where $\vc$ is the critical volume, $c$ the curvature of $U(v)$ at $\vc$, and $V$ the volume in which nucleation occurs. $P(\vc )$ is the probability density to find a bubble with volume $\vc $. Menzl~\textit{et al.} obtained $P(\vc )/V$ by umbrella sampling simulations, but they also showed that a reasonable estimate can be obtained from the free energy given by classical nucleation theory (CNT):
\beq
U(v) = p v + \gamma (36 \pi v^2)^{1/3}\, .
\label{eq:U}
\eeq
They wrote:
\beq
\frac{P(\vc )}{V} = \tilde{\rho}_0 \exp \left( - \beta U(\vc ) \right) \, .
\label{eq:Pvc}
\eeq
The choice of the prefactor $\tilde{\rho}_0$ is a delicate matter, as there is no prescription from CNT. Among various possible choices, Menzl~\textit{et al.} used $\tilde{\rho}_0 = \rL \rV$, which gives a value only 4.5 times lower than the correct value obtained from umbrella sampling simulations.

In the present work, the situation is simpler, because we are now dealing with an equilibrium system in the canonical ensemble, and $\tilde{\rho}_0 = 1/Z$ where $Z$ is the partition function. Introducing
\beq
I(a,b) = \int_a^b \exp \left[-\beta\Delta F(v) \right] \,\mathrm{d}v \, ,
\label{eq:Iab}
\eeq
we have $Z=I(0,V)$. We define the partial partition functions as $Z_\mathrm{liq}=I(0,\vc)$ and $Z_\mathrm{bub}=I(\vc,V)$ on the liquid and on the bubble side, respectively. We give here their approximate expressions. With $F_0 = V/(2 \kappa)$, $\Delta F(v)=F_0 \,\phi(x=v/V)$, and
\beq
I(a,b) = \int_a^b \exp \left[-\beta F_0 \phi(x) \right] \,\mathrm{d}v = V \int_{a/V}^{b/V} \exp \left[-\beta F_0 \phi(x) \right] \,\mathrm{d}x\, .
\label{eq:Iab2}
\eeq
We introduce $\xbub=\vbub/V$, the location of the minimum of $\phi(x)$, and the non-dimensional curvature $\hat{c}=\partial^2\phi/\partial x^2$. With $\hat{c}_\mathrm{bub}=\hat{c}(\xbub)$, the partial partition functions are:
\begin{widetext}
\begin{eqnarray}
Z_\mathrm{liq} & = & I(0,\vc) \simeq V \int_{0}^{\xc} \exp \left[-9 \beta F_0 \epsilon x^{2/3} \right] \,\mathrm{d}x \simeq \frac{3}{2} \frac{V}{(9\beta F_0 \epsilon)^{3/2}} \int_{0}^{+\infty} \sqrt{y} \, e^{-y} \,\mathrm{d}y =\frac{1}{8} \left(\frac{\kB T}{\gamma}\right)^{3/2}\, ,\label{eq:Zliq}\\
Z_\mathrm{bub}&=&I(\vc,V) \simeq V \int_{\xbub}^{1} \exp \left\{ -\beta F_0 \left[ \phi (\xbub) + \frac{\hat{c}_\mathrm{bub}}{2} (x-\xbub )^2 \right] \right\} \,\mathrm{d}x \\
&\simeq& V \exp \left[ -\beta F_0 \phi (\xbub) \right] \int_{-\infty}^{+\infty} \exp \left[ -\beta F_0 \frac{\hat{c}_\mathrm{bub}}{2} y^2 \right] \,\mathrm{d}y =\sqrt{4\pi\frac{\kB T V \kappa}{\hat{c}_\mathrm{bub}}} \exp \left[ -\beta \Delta F( \vbub ) \right] \, .\label{eq:Zbub}
\end{eqnarray}
\end{widetext}
The total partition function follows from $Z=Z_\mathrm{liq}+Z_\mathrm{bub}$.

Let $\pliq$ and $\pbub$ be the probabilities to find the system on the liquid side and on the bubble side of the free-energy barrier, respectively. We have:
\beq
\pliq = \frac{Z_\mathrm{liq}}{Z}, \quad \pbub = \frac{Z_\mathrm{bub}}{Z}, \quad\mathrm{and}\quad \frac{\pbub}{\pliq}=\frac{Z_\mathrm{bub}}{Z_\mathrm{liq}} \, .
\label{eq:pliq}
\eeq

In our model, the rate $J$ at which the system jumps over the barrier thus writes:
\beq
J = \frac{\sqrt{-\hat{c}_\mathrm{c}} D(\vc)}{\sqrt{4\pi \kB T V \kappa}} \frac{1}{Z V} \exp \left[-\beta\Delta F(\vc) \right] \, ,
\label{eq:J2}
\eeq
where $\hat{c}_\mathrm{c}=\hat{c}(\xc)$.

The only ingredient still needed to determine $J$ from Eq.~\ref{eq:J} is $D(\vc )$. Again, this can be obtained from simulations, or estimated based on a theory of the bubble dynamics in the presence of thermal fluctuations. In hydrodynamics, the Rayleigh-Plesset equation (RPE) describes the evolution of a spherical bubble~\cite{Brennen_cavitation_1995}. Using the bubble volume as the variable, the RPE writes:
\beq
\rL \ddot{v} - \frac{\rL \dot{v}^2}{6 v} = 4 \pi \left( \frac{3v}{4\pi} \right)^{1/3} \left[p_\mathrm{in} - p - 2 \gamma \left( \frac{4\pi}{3v} \right)^{1/3} -\frac{4}{3} \eta \dot{v}\right] \, ,
\label{eq:RPE}
\eeq
where $p_\mrm{in}$ is the pressure inside the bubble, and $\eta$ is the shear viscosity. In the overdamped limit, and neglecting $p_\mrm{in}$, the RPE gives:
\beq
\dot{v}=-\frac{3v}{4\eta} \left[ p + 2\gamma \left( \frac{3v}{4\pi} \right)^{1/3} \right] = -\frac{1}{\Gamma (v)} \frac{\mrm{d}U}{\mrm{d}v} \, .
\label{eq:vdot}
\eeq
This corresponds to the asymptotic velocity of an object moving in the potential $U(v)$ with a friction coefficient $\Gamma (v) = 4 \eta/(3v) $. A Langevin like random force is introduced to account for thermal fluctuations. Like in the classical case of Brownian motion, the diffusion coefficient follows from the fluctuation-dissipation theorem:
\beq
D(v)= \frac{\kB T}{\Gamma (v)}= \frac{3\kB T v}{4 \eta}\, .
\label{eq:D}
\eeq

In our case, the bubble evolves in a confined space. Vincent and Marmottant~\cite{Vincent_statics_2017} have shown that the RPE is modified by an additional inertial term, a factor in the effective mass of the bubble, and by the use of the appropriate $U(v)$ (in their case they considered the free energy for a compressible liquid and the elasticity of the surrounding medium). We note that confinement does not modify the dissipation, as the only viscous contribution to the RPE comes from the dynamic boundary condition at the bubble surface~\cite{Brennen_cavitation_1995}. Therefore, in the overdamped limit, Eq.~\ref{eq:vdot} remains unchanged, and we will use Eq.~\ref{eq:D} to compute  the value $D(\vc)$ required to obtain the rate with Eq.~\ref{eq:J}. 

At equilibrium, the system obeys detailed balance:
\beq
JV = \kliq \pliq = \kbub \pbub \, ,
\label{eq:JV}
\eeq
where $\kliq$ and $\kbub$ are the rates at which barrier crossing occurs from the liquid side and from the bubble side, respectively. We now deduce the corresponding mean first passage times (MFPTs), $\tauliq$ and $\taubub$. For the transition from liquid to bubble:
\begin{eqnarray}
\tauliq &=& \frac{1}{\kliq} = \frac{\sqrt{4\pi \kB T V\kappa}}{\sqrt{-\hat{c}_\mathrm{c}} D(\vc)} \Zliq \exp \left[\beta\Delta F(\vc) \right]\\
&=&\frac{\sqrt{\pi}}{3} \frac{\eta\kappa}{\xc} \frac{\kb T}{\sqrt{(-\hat{c}_\mathrm{c}) V \kappa \gamma^3}} \exp \left[\beta\Delta F(\vc) \right]\, ,\label{eq:tauliq}
\end{eqnarray}
while for the transition from bubble to liquid:
\begin{eqnarray}
\taubub &=& \frac{1}{\kbub} = \frac{\sqrt{4\pi \kB T V\kappa}}{\sqrt{-\hat{c}_\mathrm{c}} D(\vc)} \Zbub \exp \left[\beta\Delta F(\vc) \right]\\
&=&\frac{16 \pi}{3} \frac{\eta\kappa}{\xc} \frac{1}{\sqrt{(-\hat{c}_\mathrm{c}) \hat{c}_\mathrm{bub}}} \exp \left\{\beta\left[\Delta F(\vc)-\Delta F(\vbub) \right]\right\}\, .\nonumber \\ 
&&\quad\label{eq:taubub}
\end{eqnarray}
Note that, at the binodal, $\xc$ can be approximated using Eq.~\ref{eq:xcphicapprox}, and, similarly, $\hat{c}_\mathrm{c}\simeq -2/{A_x}^{4/3}\simeq 7.6$ and $\hat{c}_\mathrm{bub}\simeq 4/3$.

The ratio of the MFPTs satisfies:
\beq
\frac{\tau_\mathrm{bub}}{\tau_\mathrm{liq}} = \frac{\pbub}{\pliq} = \frac{Z_\mathrm{bub}}{Z_\mathrm{liq}} \, .
\label{eq:detbal}
\eeq
Note that, even though at the bubble binodal the two minima of the free energy have equal height, $\tau_\mathrm{bub} \neq \tau_\mathrm{liq}$ because $Z_\mathrm{bub} \neq Z_\mathrm{liq}$.

\subsection{Simulations\label{sec:simuls}}

To support the presented theoretical approach, we perform computer simulations of a three dimensional Lennard-Jones (LJ) fluid whose phase behavior is well established. Previous simulations have already discussed the stability and characteristics of bubbles in confined LJ systems~\cite{Park_molecular_2001,Neimark_birth_2005,Neimark_phase_2006,MacDowell_nucleation_2006}. Particles of mass $m$ interact via a LJ potential with energy and size parameters $\epsilon_\mathrm{LJ}$, and $\sigma_\mathrm{LJ}$, respectively. The potential was truncated with a cutoff of $2.5\sigma_\mathrm{LJ}$. All results are presented in reduced Lennard-Jones units. Table~\ref{table1} gives literature values at $T=0.75$ for the liquid density at saturated vapor pressure $\rho_\mathrm{L}^\infty$, the surface tension $\gamma$, the isothermal compressibility $\kappa$, the Berthelot-Laplace length $\lambda$, and the shear viscosity $\eta$.

\begin{table}[bh!]
\centering
\begin{ruledtabular}
\caption{Properties of the Lennard-Jones liquid at $T=0.75$ and saturated vapor pressure. Values between parentheses give the uncertainty on the last digit.
\label{table1}
}
\begin{tabular}{cccc}
Parameter & Units   & Value &   Ref.  \\
$\rho_\mathrm{L}^\infty$ & ${\sigma_\mathrm{LJ}}^{-3}$& 0.822(2) &\onlinecite{Chen2001}\\
$\gamma$ & $\epsilon_\mathrm{LJ}\,{\sigma_\mathrm{LJ}}^{-2}$ & 1.04(4) & \onlinecite{Stephan2019}    \\
$\kappa$ & ${\epsilon_\mathrm{LJ}}^{-1}{\sigma_\mathrm{LJ}}^3$ & 0.091(3) & \onlinecite{Lotfi1992}     \\
$\lambda$ & ${\sigma_\mathrm{LJ}}$ & 0.063(5)  &   \\
$\eta$  &   $\sqrt{m \epsilon_{LJ}}\,{\sigma_\mathrm{LJ}}^{-2}$  & 0.81\footnote{This approximate value is obtained by linear extrapolation to $T=0.75$ of data at higher temperature and $\rho=0.7$} & \onlinecite{Meier_transport_2004}
\end{tabular}
\end{ruledtabular}
\end{table}

\begin{figure}[th]
\includegraphics[width=0.7\columnwidth]{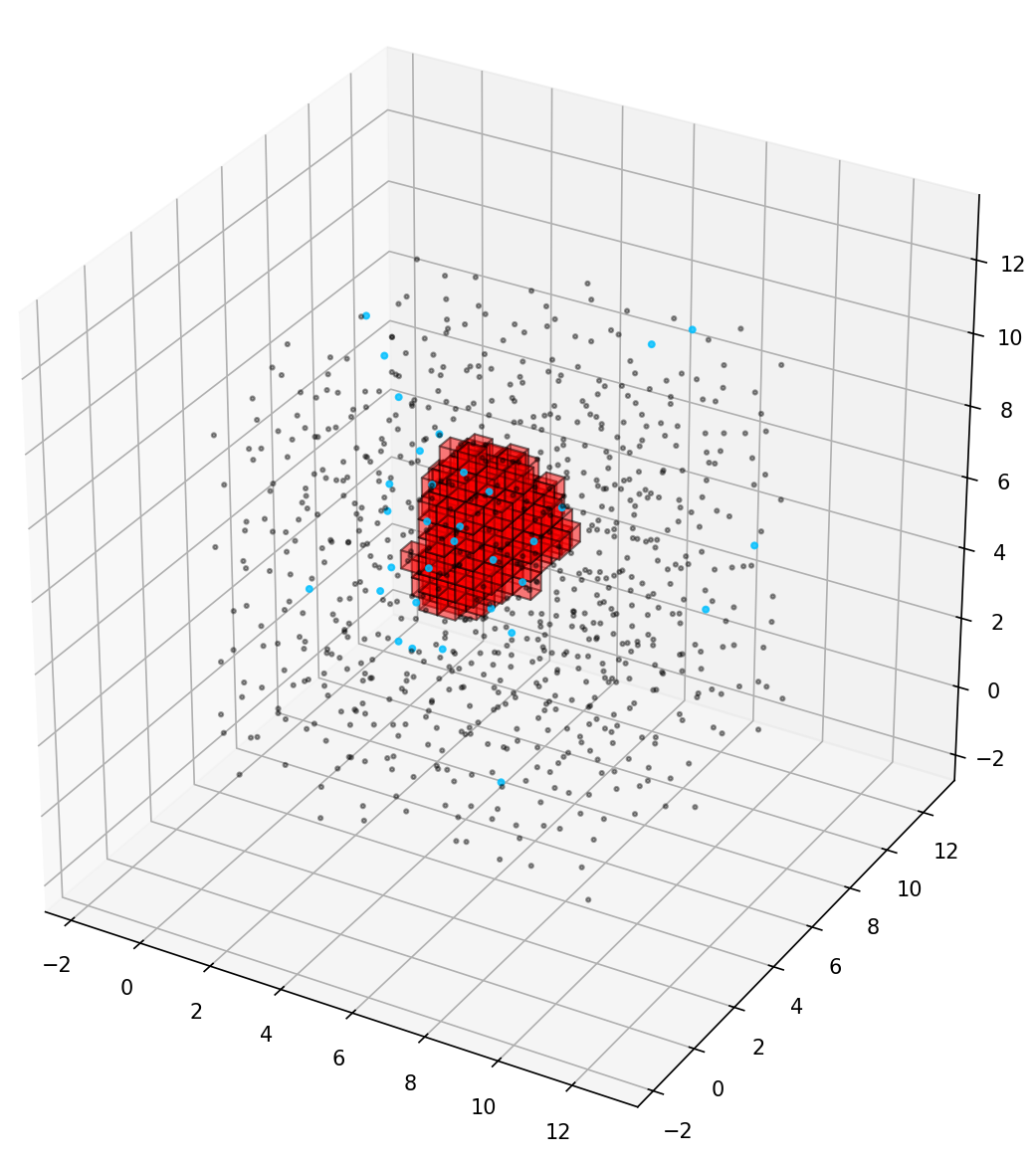}
\caption{Snapshot of a typical bubble created following the protocol reported in the text. The largest bubble is depicted as red while the liquid-like and non liquid-like LJ particles are shown in blue and light blue, respectively. Note that the size of the LJ particles was decreased for a better view of the bubble.\label{fig:bubble}}
\end{figure}

Molecular dynamics simulations are performed using the 
LAMMPS open source package.\cite{Thompson_lammps_2022} The system was initialized in a box with periodic boundary conditions containing $N = 800$ particles. At first, a nanobubble was generated by removing atoms from a spherical region at the center of the box, corresponding to approximately 15\% of the total simulation volume (see Fig.~\ref{fig:bubble}). To maintain the target global density, an equivalent number of particles was re-inserted into the region outside the bubble. Subsequently, the system energy was minimized over 1,000 iterations. Production runs were conducted in the $NVT$ ensemble using a Nosé-Hoover thermostat~\cite{Nose1984, Hoover1985, Martyna1992} at $T=0.75$  with a reduced timestep of $\Delta t = 0.005$. To ensure proper statistical sampling, the simulation length was $6\,10^6$ total integration steps. In order to avoid potential errors due to the initial bubble inserted, the analysis was performed ignoring the first 5000 steps.

\begin{figure}[t]
\includegraphics[width=0.9\columnwidth]{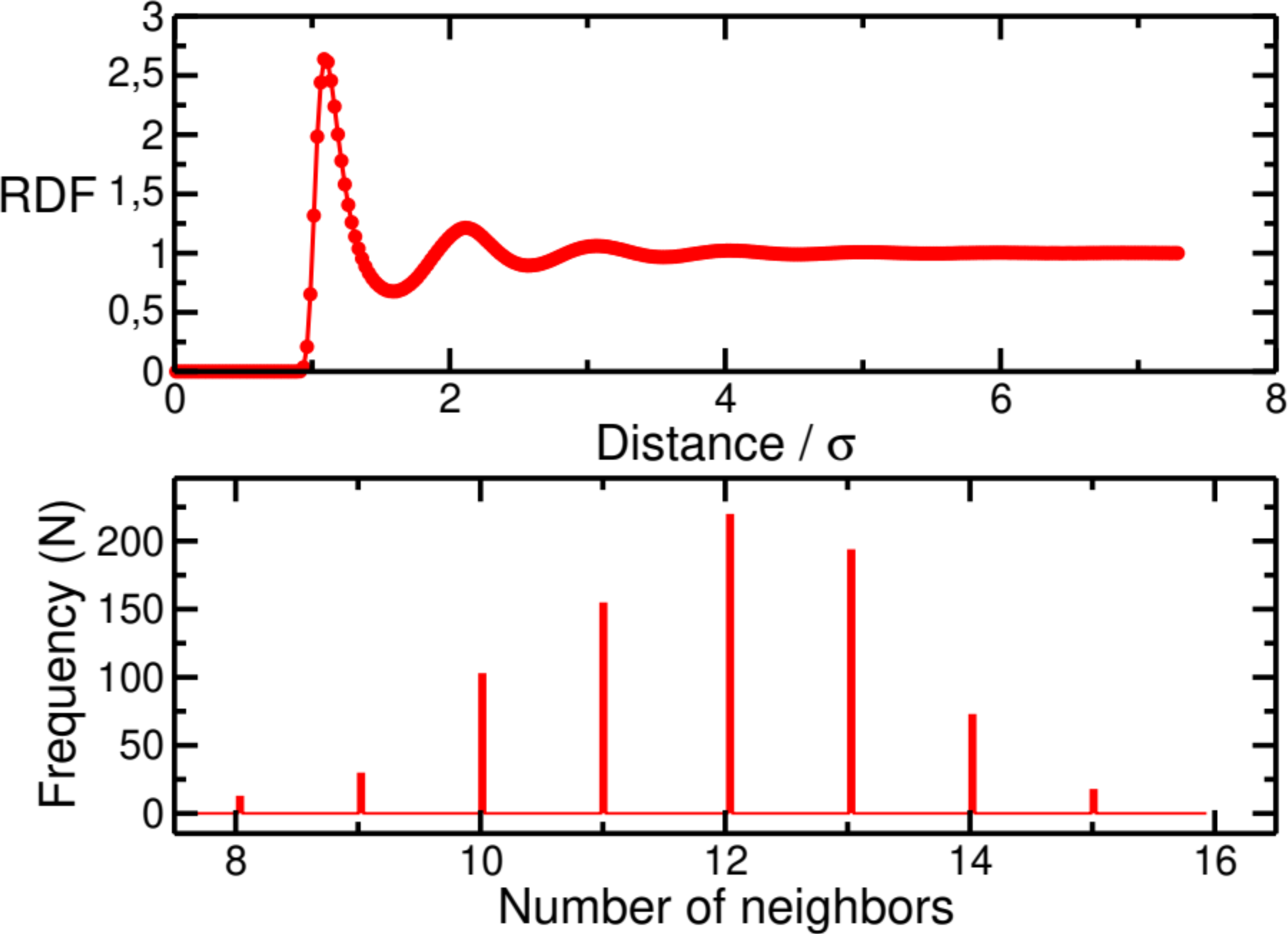}
\caption{Top: radial distribution function $g(r)$ at $T=0.75$ and $\rho=0.73$. Bottom: histogram of the number of neighbors within a sphere of radius $r_\mathrm{C} = 1.6 \sigma$, corresponding to the first minimum of the radial distribution function $g(r)$.\label{fig:rdf}}
\end{figure}

In order to identify the largest bubble, we use a procedure similar to the one used in Ref.~\onlinecite{Wang_homogeneous_2009}:
\begin{enumerate}
\item We use Stillinger’s cluster criterion~\cite{Stillinger_rigorous_1963,tenWolde_numerical_1999} to identify the nearest neighbors of each particle: according to our definition, two particles are neighbors if their distance is less than a cut-off $r_\mathrm{C} = 1.6 \sigma$, corresponding to the first minimum of the radial distribution function $g(r)$ at $T=0.75$ and $\rho=0.73$.
\item Using this criterion, we compute the nearest neighbors probability distribution function in the liquid at $T=0.75$ and $\rho=0.73$, from which we decide  to define particles having at least 8 neighbors as liquid-like (see Fig.~\ref{fig:rdf}).
\item Following Ref.~\onlinecite{Wang_pore_2005}, we build a three-dimensional grid with cubic cells of side $\sigma/2$ and label the grid cells according to the nature of the particles in the cell, either liquid or non-liquid; if a cell is empty, it is labelled as non-liquid.
\item Having detected all non-liquid cells, we cluster neighbouring vapor cells which will form all bubbles in the system and choose the volume of the largest of those as our order parameter.
\end{enumerate}

For each frame, this procedure yields the volume of the largest bubble $v$. The $v$ values obtained during a run are binned to generate a histogram and obtain the probability density $P(v)$. To efficiently sample the $v$ values with good statistical accuracy, we use a type of adaptive binning. We sort the $v$ values in ascending order, and use bins of variable width, with a fixed number of counts $N_\mathrm{counts}$ in each bin, for all $v$ less than a threshold slightly higher than the critical bubble volume. For higher $v$ values, we also use bins of variable width, but with a higher fixed number of counts $5 N_\mathrm{counts}$ in each. We used $N_\mathrm{counts}$ ranging from 340 to 3200 for densities $\rho \sigma^3$ ranging from 0.635 to 0.645. The probability density $P(v)$, with $v$ the volume at the center of the bin, follows as $N_\mathrm{bin}/(N_\mathrm{f} \, w)$, where $N_\mathrm{bin}=N_\mathrm{counts}$ or $5 N_\mathrm{counts}$, respectively, $N_\mathrm{f}$ is the total number of frames, and $w$ the width of the current bin. Finally, the free-energy is $U(v)=-\kB T \ln [v_0 P(v)]$, where $v_0$ is a normalization constant set to get $U=0$ for the first bin.

\section{Results\label{sec:results}}

\subsection{Bubble collapse\label{sec:collapse}}

In this section, we investigate the conditions at which bubble collapse will occur due to thermal fluctuations in a small cavity. We consider the case of pure water, and illustrate the results for two temperatures, 20 and \qty{300}{\degreeCelsius}.

\begin{figure}[t]
\centering
\includegraphics[width=0.95\columnwidth]{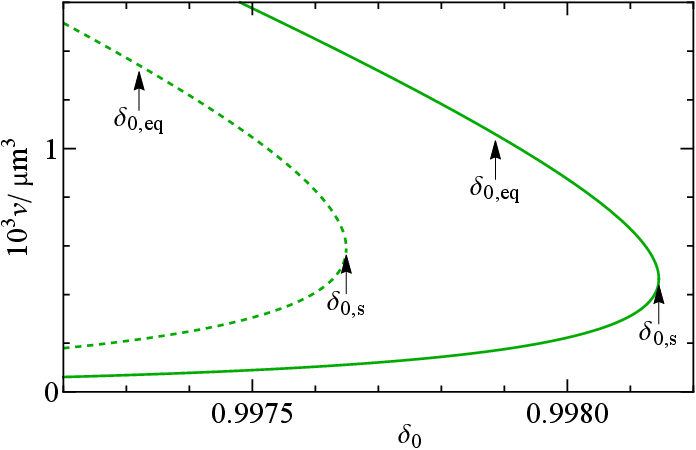}
\caption{Characteristic bubble volumes as a function of reduced average density for pure water. The metastable and critical bubble volumes correspond to the upper and lower portion of the curves, respectively. Here the cavity volume is $V=\qty{1}{\mu m^3}$, and the temperature is $T=\qty{20}{\degreeCelsius}$ (solid curve) or \qty{300}{\degreeCelsius} (dotted curve). The corresponding locations of the binodals and spinodals are indicated with arrows.
\label{fig:V}
}
\end{figure}

Figure~\ref{fig:V} shows the equilibrium and critical bubble volumes, $\ve$ and $\vc$, respectively, as a function of the reduced average density $\dav$, for a cavity with volume $V=\qty{1}{\mu m^3}$. By definition, the two branches $\ve$ and $\vc$ meet at the bubble spinodal, $\ds$. $\de$ and $\ds$ decrease with increasing temperature. In regime (ii), where the bubble is metastable, $\vc > 1.7\,10^{-4}\unit{\mu m^3}$, which corresponds to a critical bubble radius above \qty{34}{nm}. This justifies the use of macroscopic thermodynamics to describe the system. Only for $V=10^{-6}\,\unit{\mu m^3}$ and $T=\qty{20}{\degreeCelsius}$, the critical radius reaches around \qty{1}{nm}, which is comparable to the thickness of the liquid-vapor interface~\cite{Caupin_liquidvapor_2005}, and makes the macroscopic description approximate.

\begin{figure}[b]
\centering
\includegraphics[width=0.95\columnwidth]{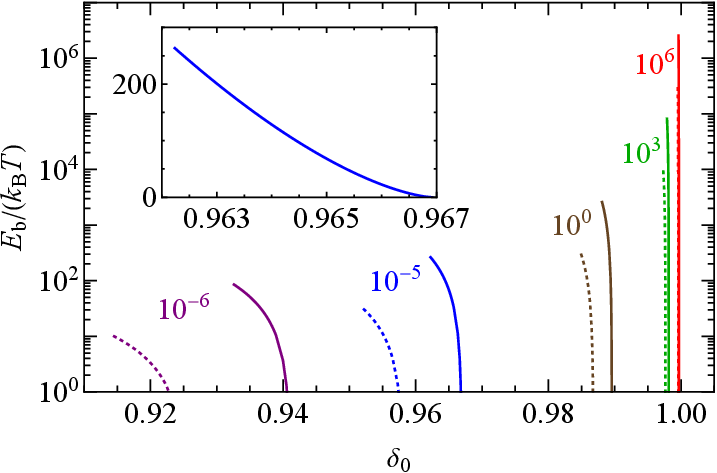}
\caption{Energy barrier for bubble collapse as a function of reduced average density for various cavity volumes (as given by the labels in \unit{\mu m^3}), at $T=\qty{20}{\degreeCelsius}$ (solid curves) or $\qty{300}{\degreeCelsius}$ (dotted curves). Each curve starts at the corresponding $\de$. The inset shows the case $V=10^{-5}\unit{\mu m^3}$, $T=\qty{20}{\degreeCelsius}$, on a linear scale.
\label{fig:Eb}
}
\end{figure}
\begin{figure}[t]
\centering\includegraphics[width=0.95\columnwidth]{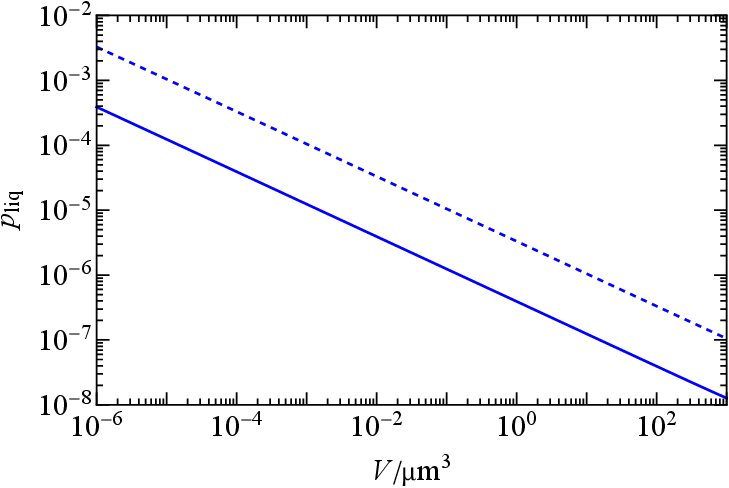}
\caption{Probability $\pliq$ of observing the system on the liquid side (Eq.~\ref{eq:pliq}) as a function of cavity volume, at the bubble binodal and $T=\qty{20}{\degreeCelsius}$ (solid curve) or $\qty{300}{\degreeCelsius}$ (dotted curve).
\label{fig:pliq}
}
\end{figure}
\begin{figure}[b]
\centering
\includegraphics[width=0.95\columnwidth]{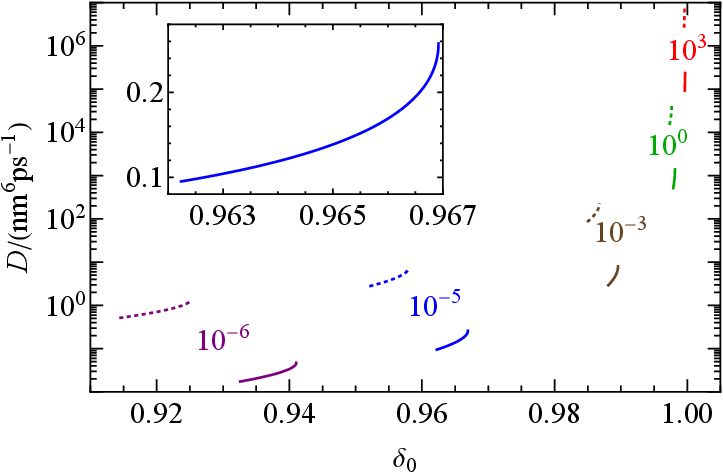}
\caption{Diffusion coefficient at the top of the barrier, $D(\vc )$ (Eq.~\ref{eq:D}), as a function of reduced average density for various cavity volumes (as given by the labels in \unit{\mu m^3}), at $T=\qty{20}{\degreeCelsius}$ (solid curves) or $\qty{300}{\degreeCelsius}$ (dotted curves). Each curve starts at the corresponding $\de$ and ends at $\ds$.
\label{fig:D}
}
\end{figure}
\begin{figure}[h!]
\centering
\includegraphics[width=0.95\columnwidth]{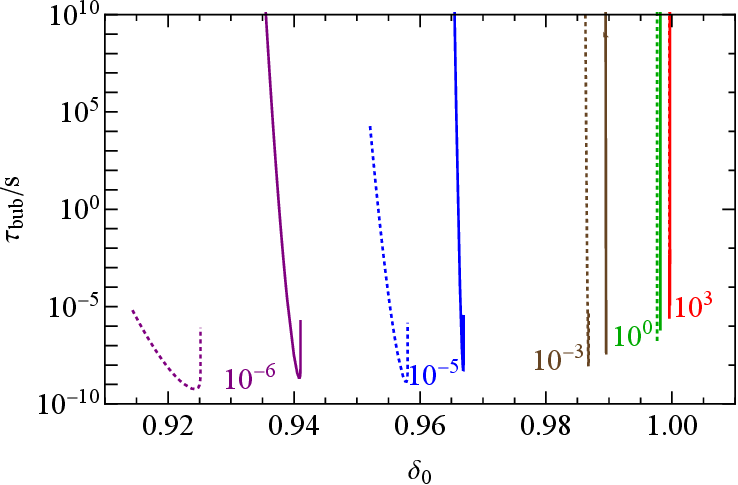}
\caption{MFPT from the bubble side $\taubub$ (Eq.~\ref{eq:taubub}) as a function of reduced average density for various cavity volumes (as given by the labels in \unit{\mu m^3}), at $T=\qty{20}{\degreeCelsius}$ (solid curves) or $\qty{300}{\degreeCelsius}$ (dotted curves). For each curves, the right endpoints correspond to $\ds$. For the two smallest cavity volumes, the left endpoints correspond to $\de$.
\label{fig:tau}
}
\end{figure}

In the following, we focus on regime (ii) (see Fig.~\ref{fig:phi}) where the bubble is metastable. It is separated from the stable fully liquid state by an energy barrier $\Eb$, defined as the difference between the local maximum in free energy (corresponding to the critical bubble) and the local mininum at large volume (corresponding to the metastable bubble). Figure~\ref{fig:Eb} shows $\Eb$ obtained from Eq.~\ref{eq:phi} in units of $\kB T$ as a function of the reduced average density $\dav$. The energy range covers several orders of magnitude. Indeed, $\Eb$ vanishes at $\ds$, but it can reach huge values at $\de$, depending on the cavity volume. A rough estimate of the barrier that could lead to bubble collapse is a few tens of $\kB T$. For cavity volumes $V \geq \qty{1}{\mu m^3}$ relevant to geosciences, this will occur only close to $\ds$. In contrast, for smaller cavity volumes, this may occur for $\dav$ significantly less than $\ds$. In the following, we study this quantitatively based on the approach by Menzl~{\etal}, as presented in Section~\ref{sec:rate}. Before doing so, we show on Fig.~\ref{fig:pliq} the probability $\pliq$ of observing the system on the liquid side at the bubble binodal. Note that this assumes the system to be at equilibrium, sampling all possible configurations, which may prove impossible on a reasonable timescale when $\Eb$ is large. Nevertheless, Fig.~\ref{fig:pliq} shows that the calculated $\pliq$ always remain small (and consequently scales as $1/\sqrt{V}$, see Eqs.~\ref{eq:Zliq} and \ref{eq:Zbub}). This points toward an ambiguity in defining the bubble binodal. The usual convention is that it corresponds to the conditions at which the two minima in the free energy have equal height. This ensures that the states corresponding exactly to the minima have equal probability. However, the probabilities of being on either side of the barrier are very different, due to the overwhelming number of states located on the bubble side.

Finally, we  compute the diffusion coefficient at the top of the barrier using Eq.~\ref{eq:D}. The results are shown on Fig.~\ref{fig:D}. They are very sensitive to the cavity volume $V$, because of $D\propto \vc \propto V^{3/4}$. Only for $V=10^{-6}\,\unit{\mu m^3}$, for which $\vc$ reaches the \unit{nm^3} range, one finds $D$ in the range of $10^{-2}\,\unit{nm^3 ps^{-1}}$; this is close to the values reported by Menzl~{\etal} for cavitation\cite{Menzl_molecular_2016}, which also involves nanoscopic critical bubbles.

\begin{figure}[t]
\centering
\includegraphics[width=0.95\columnwidth]{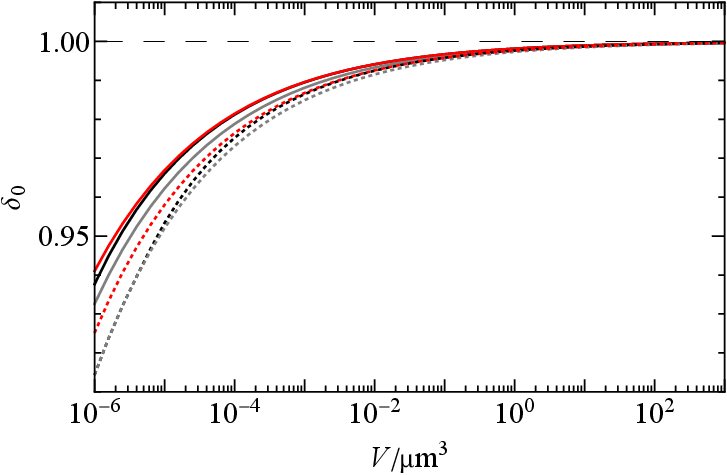}
\includegraphics[width=0.95\columnwidth]{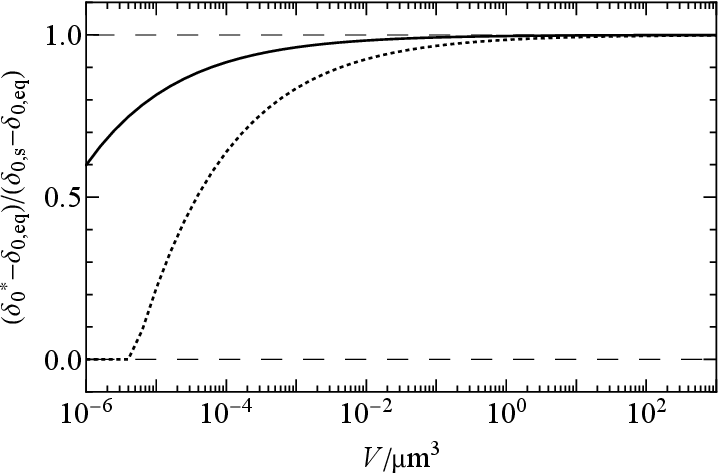}
\caption{Top: characteristic reduced average densities as a function of cavity volume; from bottom to top: $\de$ (gray), $\done$ at which $\taubub=\qty{1}{s}$ (black), and $\ds$ (red). Bottom: relative location of the reduced density $\done$ at which $\taubub=\qty{1}{s}$ as a function of cavity volume. Here the temperature is $T=\qty{20}{\degreeCelsius}$ (solid curves) or $\qty{300}{\degreeCelsius}$ (dotted curves).
\label{fig:dens}
}
\end{figure}

We now compute the MFPT from the bubble side using Eq.~\ref{eq:taubub}. The result is shown on Fig.~\ref{fig:tau}. For a given cavity volume at a given temperature, the bubble MFPT decreases with increasing density. Note that, according to Eq.~\ref{eq:taubub}, $\taubub$ diverges at $\ds$ because of the vanishing curvature $c$. This is responsible for the minima reached in Fig.~\ref{fig:tau}, although the end of the curves is not captured due to limited numerical accuracy. However, when approaching the bubble spinodal at $\ds$, the energy barrier becomes small and Kramers' result and our approximation for $\Zbub$ (Eq.~\ref{eq:Zbub}) become inaccurate: $\taubub$ should approach zero at $\ds$. For large cavity volumes, $\taubub$ increases extremely rapidly when $\dav$ decreases away from $\ds$. This is in line with the comment made above about the need to be very close to $\ds$ to keep barriers low enough to be overcome.

Of particular interest to experiments are the conditions at which bubble collapse will become observable in a reasonable time, e.g. \qty{1}{s}. Figure~\ref{fig:dens} shows $\done$, the $\dav$ value at which this happens, as a function of cavity volume. By definition, $\done$ lies in between $\de$ and $\ds$, also shown in Fig.~\ref{fig:dens}. For large cavity volumes $V$, $\done$ is in fact extremely close to $\ds$, which is required to reach $\Eb$ values which can be overcome by thermal fluctuations. Only for small $V$ and high $T$ can $\done$ come closer to $\de$. 

\begin{figure}[t]
\centering
\includegraphics[width=0.95\columnwidth]{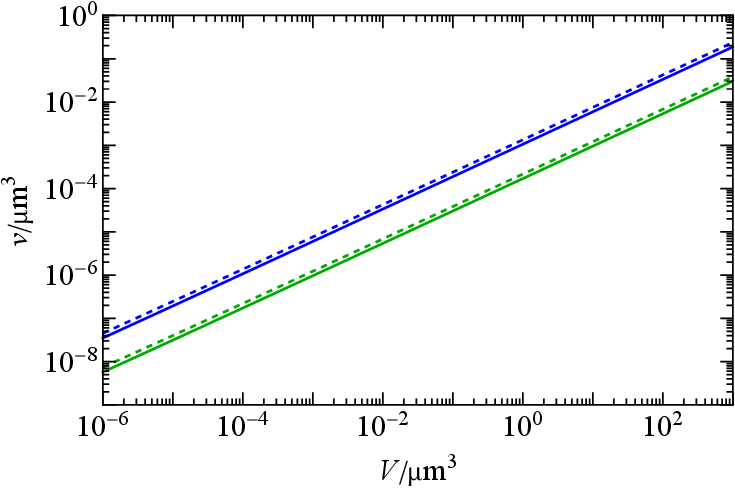}
\includegraphics[width=0.95\columnwidth]{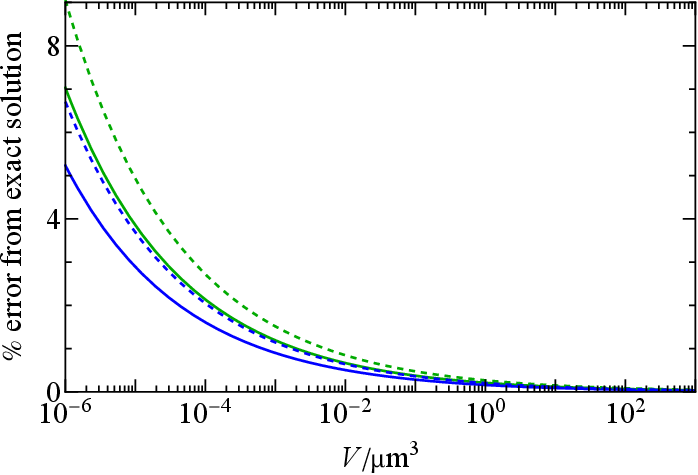}
\caption{Top: characteristic volumes at the bubble binodal a function of cavity volume, at $T=\qty{20}{\degreeCelsius}$ (solid curves) or $\qty{300}{\degreeCelsius}$ (dotted curves). For each temperature, the top blue curve gives $\ve$, and the bottom green curve $\vc$. Bottom: percent error on the characteristic volumes at the bubble binodal as a function of cavity volume, when using the analytic expression Eq.~\ref{eq:xcphicapprox}, at $T=\qty{20}{\degreeCelsius}$ (solid curves) or $\qty{300}{\degreeCelsius}$ (dotted curves). For each temperature, the bottom blue curve gives the error on $\ve$, and the top green curve that on $\vc$.
\label{fig:Vflip}
}
\end{figure}

These observations are more clearly illustrated on Fig.~\ref{fig:dens} (bottom), which shows the normalized distance between $\de$ and $\ds$, defined as $\donerel=(\done-\de)/(\ds-\de)$. At \qty{20}{\degreeCelsius}, for cavity volumes $V \geq \qty{1}{\mu m^3}$ relevant to geosciences, $\donerel \geq 0.993$, showing that one needs to be very close to the bubble spinodal to observed bubble collapse. Only at high temperatures and for small cavity volumes is collapse expected to occur more significantly away from the spinodal. In fact, for the smallest cavities at high temperature, the bubble lifetime is always less than \qty{1}{s}, even at the bubble binodal, causing the $\donerel$ curve at \qty{300}{\degreeCelsius} to reach 0 for $V<4\,10^{-6}\,\unit{\mu m^3}$. Indeed, for these conditions, the energy barrier at the binodal becomes less than $20 \kb T$ (see also Fig.~\ref{fig:Eb}).

\subsection{Fluctuations at the bubble binodal\label{sec:bino}}

We now turn to the case in which the energy barrier is sufficiently small to allow spontaneous thermal fluctuations to make the system oscillate between the fully liquid state, and the state with a temporarily well-defined bubble (phase flipping). We will focus on the bubble binodal, which corresponds to the conditions at which the free-energy of the homogenous liquid and the stable bubble are equal.

\begin{figure}[t]
\centering
\includegraphics[width=0.95\columnwidth]{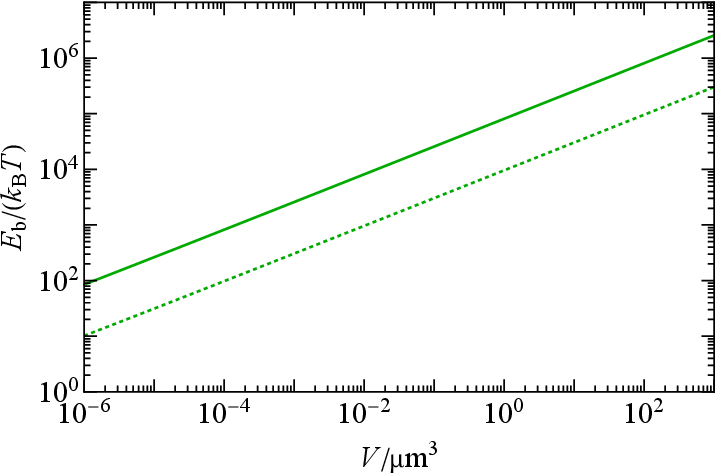}
\includegraphics[width=0.95\columnwidth]{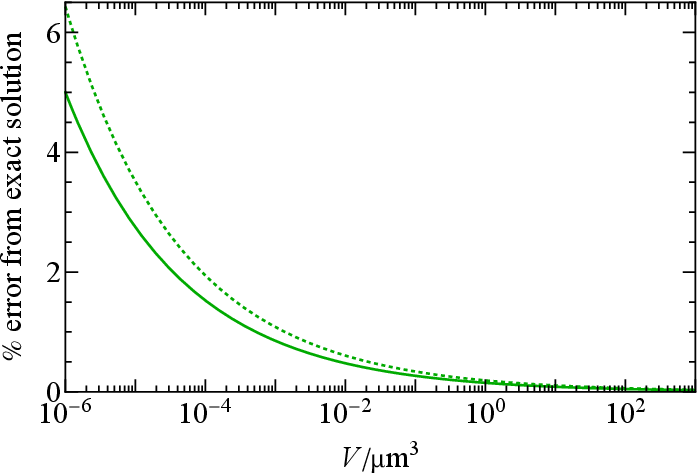}
\caption{Top: energy barrier at the bubble binodal a function of cavity volume. Bottom: percent error on the energy barrier at the bubble binodal as a function of cavity volume, when using the analytic expression Eq.~\ref{eq:xcphicapprox}. Here the temperature is $T=\qty{20}{\degreeCelsius}$ (solid curves) or $\qty{300}{\degreeCelsius}$ (dotted curves).
\label{fig:Ebflip}
}
\end{figure}

\begin{figure}[t]
\centering
\includegraphics[width=0.95\columnwidth]{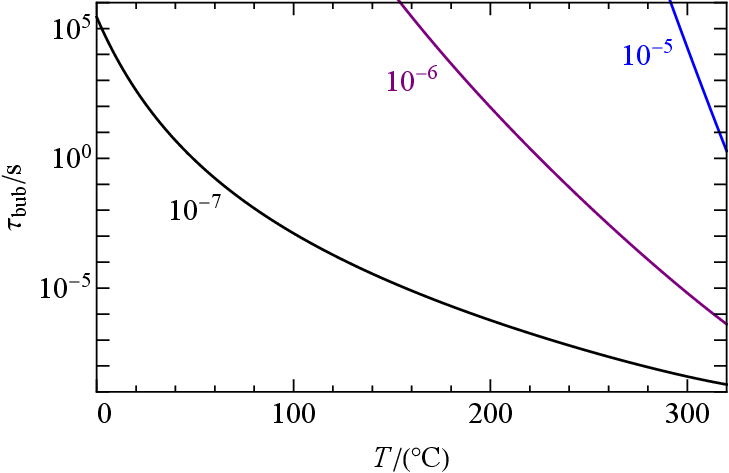}
\caption{MFPT from the bubble side as a function of temperature for various cavity volumes (as given by the labels in \unit{\mu m^3}).
\label{fig:tauflip}
}
\end{figure}

We start with a description of the characteristic parameters for the bubble binodal. Figure~\ref{fig:Vflip} (top) shows the equilibrium and critical bubble volumes as a function of cavity volume. The analytic formulas (Eq.~\ref{eq:xcphicapprox}) give excellent approximations, with a relative error always less than 9\% 
(see Fig.~\ref{fig:Vflip} (bottom)).

Figure~\ref{fig:Ebflip} (top) shows the energy barrier at the bubble binodal as a function of cavity volume. The analytic formula (Eq.~\ref{eq:xcphicapprox}) gives an excellent approximation, with a relative error always less than 7\% (see Fig.~\ref{fig:Ebflip} (bottom)). Importantly, Fig.~\ref{fig:Ebflip} shows that, as phase flipping requires moderate energy barriers, it needs small cavity volumes and high temperatures to be observable.

To quantitatively assess the possibility of observing phase flipping, we compute the MFPT from the bubble side using Eq.~\ref{eq:taubub}. The results are shown as a function of temperature on Fig.~\ref{fig:tauflip} for three small cavity volumes; larger values will preclude the observation of phase flipping in a reasonable time frame. We are particularly interested in numerical simulations, which can be performed for such small cavity volumes, on a timescale up to a few \unit{\mu s}. This can exceed the bubble lifetime at high enough temperature, with the minimum required temperature becoming smaller as the cavity volume decreases. This is because the energy barrier at the binodal scales like $\sqrt{V}$ (see Eq.~\ref{eq:xcphicapprox}). To check if phase flipping indeed happens, we performed simulations of a Lennard-Jones fluid as discussed in the next section.

\subsection{Phase flipping in a Lennard-Jones fluid\label{sec:simulsLJ}}

In this section we discuss results obtained with NVT simulations of a LJ fluid as described in Section~\ref{sec:simuls}. All parameters are given in LJ units. The temperature is $T=0.75$, the number of particles is fixed to $N=800$, and the density $\rho$ is varied by changing the cavity volume $V=N/\rho$.

\begin{figure}[t]
\centering
\includegraphics[width=1.\columnwidth]{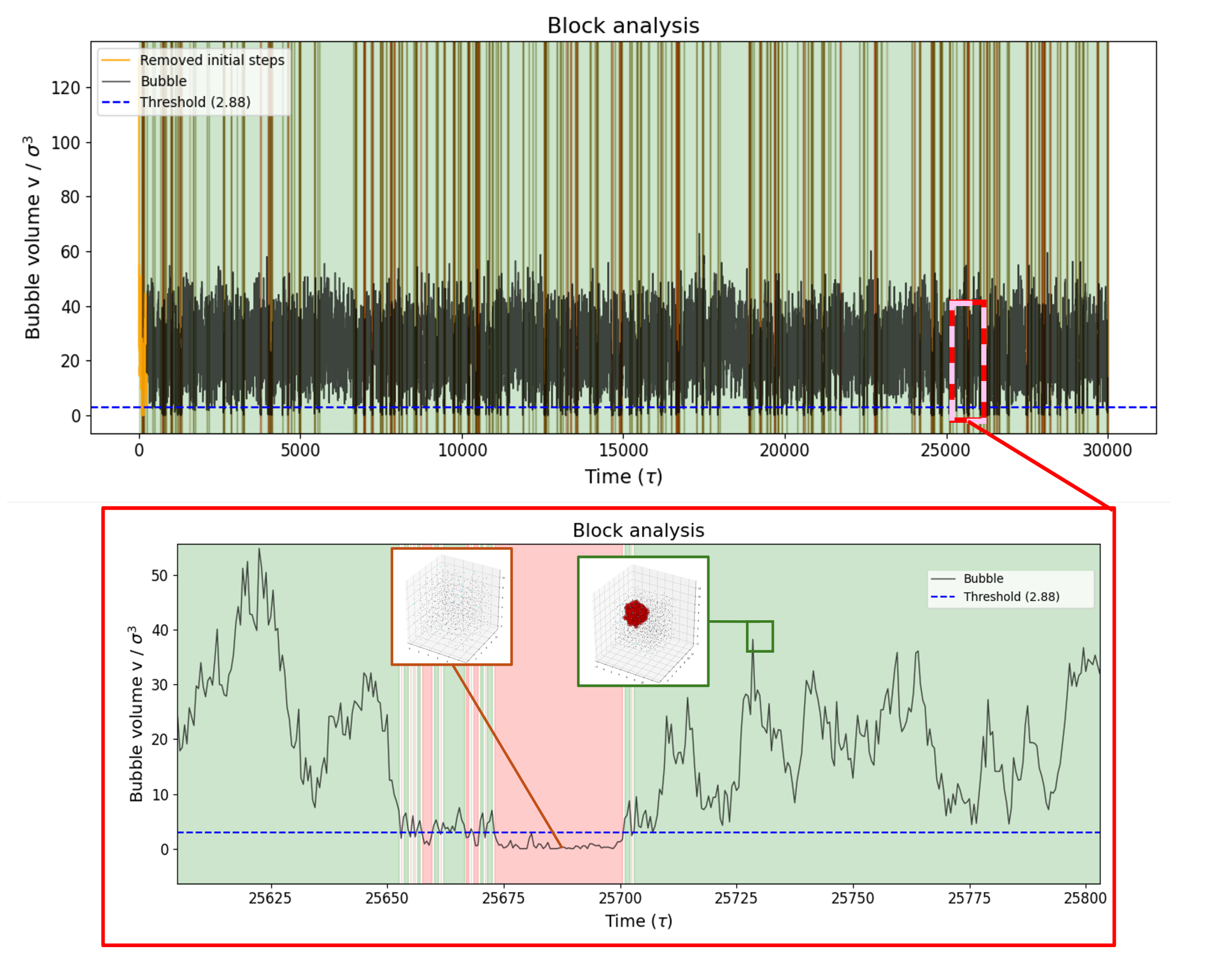}
\caption{Trajectory showing the effective bubble volume $v$ as a function of time for $T=0.75$ and $\rho=0.638$. The first 5000 steps, depicted in orange, were removed to avoid artifacts in our analysis that could have been due to the initial bubble. A threshold value on $v$ defines the portions of the trajectory on the liquid side (red) and on the bubble side (green).  The lower panel shows a close-up with illustrative snapshots.
\label{fig:blocks}
}
\end{figure}

\begin{figure}[t!]
\centering
\includegraphics[width=0.95\columnwidth]{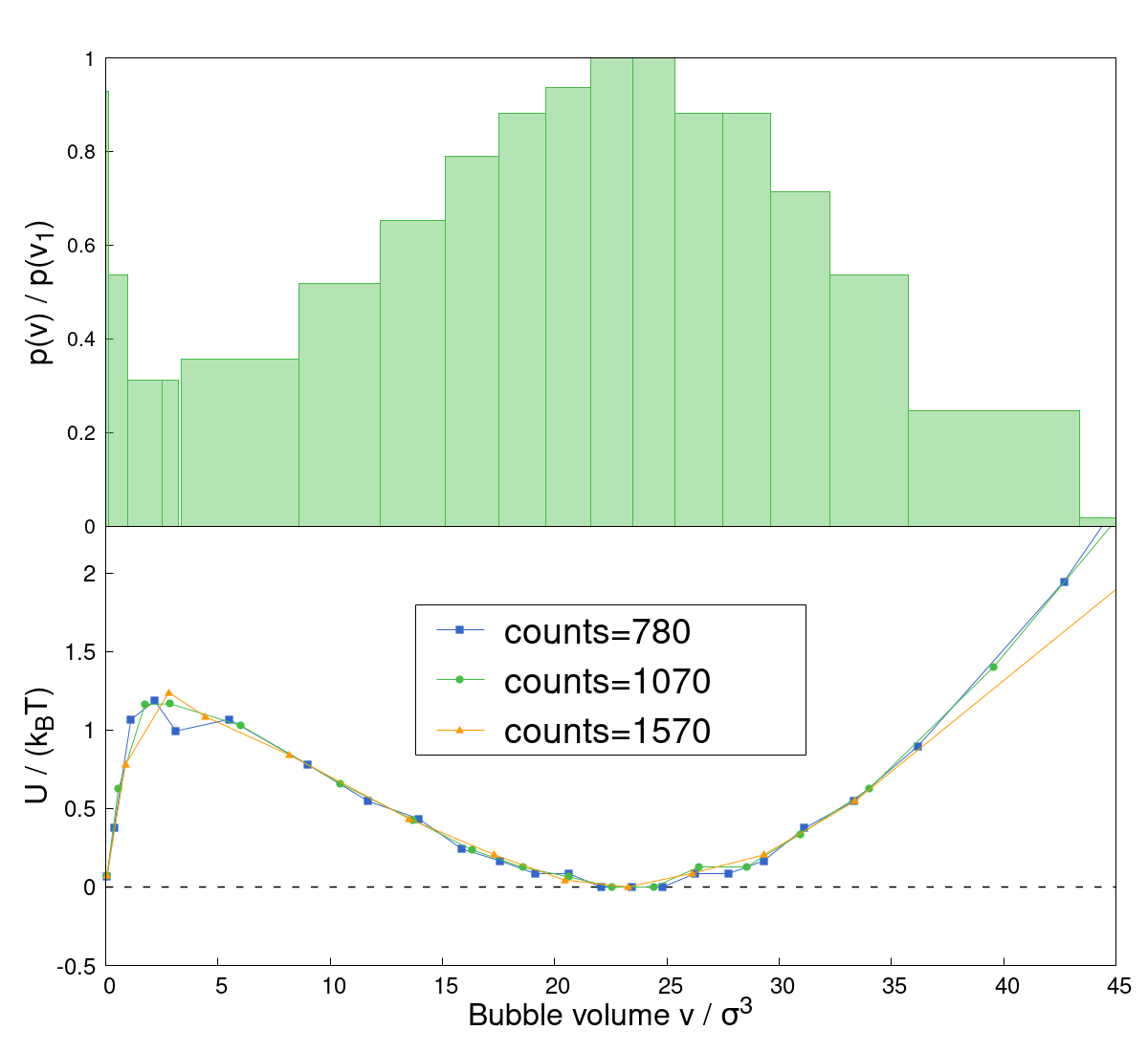}
\caption{Top: probability density  $p(v)$ of the bubble volume $v$, normalized by the value at the smallest volume $p(v_1)$, for the trajectory of Fig.~\ref{fig:blocks} at $T = 0.75$ and $\rho=0.638$. Here the bins are defined using $N_\mathrm{counts}=1070$ (see Section~\ref{sec:methods}). Bottom: corresponding free-energy $U(V)/(\kB T)$ for several choices of $N_\mathrm{counts}$. The location of the local maximum of $U(v)$ on the left sets the threshold used in Fig.~\ref{fig:blocks} to define liquid and bubble sides.
\label{fig:histo}
}
\end{figure}
\begin{figure}[t!]
\centering
\includegraphics[width=0.95\columnwidth]{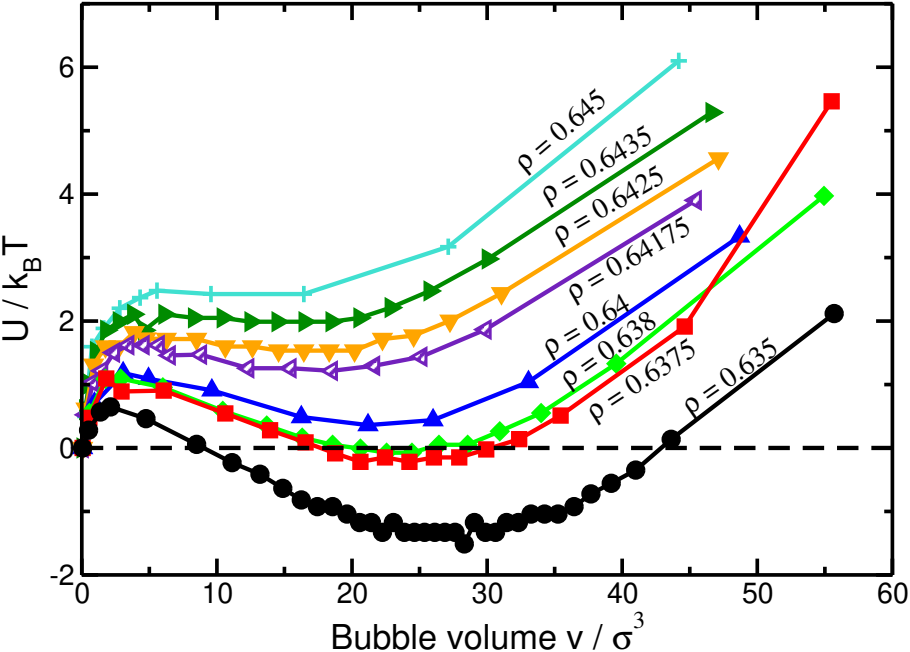}
\caption{Free-energy $U/(\kB T)$ obtained from simulations at $T=0.75$ as a function of bubble volume $v$ for several densities $\rho$.
\label{fig:LJcurves}
}
\end{figure}

Figure~\ref{fig:blocks} shows a run exhibiting phase flipping. The time is given in units of $t^\ast = \sigma_\mathrm{LJ} \sqrt{m/\epsilon_\mathrm{LJ}}$. Over the course of the run, the system oscillates between a state where the liquid is homogeneous, and a state where a relatively large bubble with fluctuating volume is present. As illustrated in Fig.~\ref{fig:histo}, the values of the bubble effective volume are binned as described in Section~\ref{sec:simuls} to generate the free-energy. The bottom panel shows that the result is robust with respect to reasonable changes of $N_\mathrm{counts}$. This procedure is repeated at several densities. Figure~\ref{fig:LJcurves} shows that, when the density is varied, the free-energy changes in a similar way to the free energy of our model (Fig.~\ref{fig:phi}). There is a range of densities, lower than the liquid density at saturated vapor pressure, in which bubbles are unstable, as predicted by the model for $\ds < \dav < \de$ (regime (i)) and observed in previous simulations~\cite{MacDowell_nucleation_2006}.

\begin{figure}[h!]
\centering
\includegraphics[width=0.95\columnwidth]{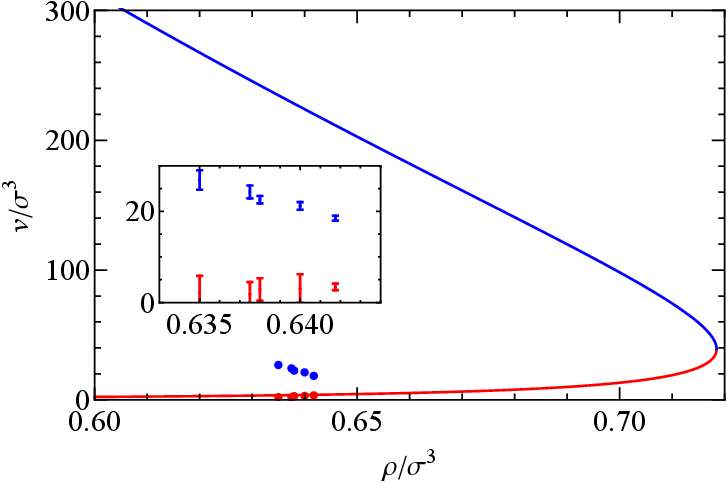}
\caption{Characteristic bubble volumes $v$ as a function of $\rho$ for a LJ system. The equilibrium and critical bubble volumes are shown in blue and red, respectively, for experiments (solid curves) and simulations (symbols). The inset shows a close-up on the simulation results.
\label{fig:Vcomp}
}
\end{figure}

We now compare the simulation results to the model predictions, which are displayed in Table~\ref{table2}. Note that the cavity volume in the model was varied as in the simulations. For regimes (ii) and (iii), where the free energy shows two minima, we report in Fig.~\ref{fig:Vcomp} the respective volumes of the critical and most probable bubble as a function of density. They meet at the bubble spinodal. Overall, the observed volumes are smaller than the model predictions, and the density of the bubble spinodal in the simulations ($\rho=0.645$) is significantly smaller than the model prediction ($\rho=0.718$). In the simulations, the bubble binodal at $T=0.75$ corresponds to $\rho\approx 0.638$, with an energy barrier around $1.1\,\kb T$, significantly lower than the model predictions, $0.704$ and $12.5\,\kb T$. Using the analysis summarized in Fig.~\ref{fig:blocks}, we obtain the MFPTs as the average values of the duration of the blocks with the system on the bubble and on the liquid side; we find $\tauliq=0.22$ and $\taubub=4.48$, respectively. These values are much smaller than the model predictions, $2.2\,10^4$ and $7.7\,10^6$, respectively. Most of the discrepancy stems from the difference in energy barrier, which contributes exponentially (see Eqs~\ref{eq:tauliq} and \ref{eq:taubub}), introducing a factor around $9\,10^4$. The agreement is better for the ratio of the two MFPTs: in the simulations, $\taubub/\tauliq=20.4$, a factor 17 less than the model prediction for $\taubub/\tauliq = \pbub/\pliq = 350$.

\begin{table}[tt]
\centering
\begin{ruledtabular}
\caption{Comparison between model predictions and simulations based on the parameters given in Table~\ref{table1} for a system of 800 particles. Values between parentheses give the uncertainty on the last digit. The last six lines are calculated at the respective density for the bubble binodal, $\rho_{0,\mathrm{eq}}=0.704$ for the model and $\rho=0.638$ for the simulations.
\label{table2}
}
\begin{tabular}{cccc}
Parameter & Units   & Model	& Simul.  \\
$\rho_{0,\mathrm{sp}}$ & ${\sigma_\mathrm{LJ}}^{-3}$ & 0.718(5) & 0.645(1) \\
$v_\mathrm{sp}$ & ${\sigma_\mathrm{LJ}}^3$ & 39.1(15) & 10(5)\\
$\rho_{0,\mathrm{eq}}$ & ${\sigma_\mathrm{LJ}}^{-3}$ & 0.704(5)     & 0.6380(2)\\
$v_\mathrm{eq}$ & ${\sigma_\mathrm{LJ}}^3$ & 89.5(34)     & 22.6(8)\\
$\epsilon$ & --- & 0.0097(5) & 0.0094(5) \\
$\Delta F (\vc ) /(\kB T)$ & --- & 12.5(10)& 1.1(2) \\
$\tauliq$ & $\sigma_\mathrm{LJ} (m/\epsilon_\mathrm{LJ})^{1/2}$    & $2.2\,10^4$& $0.22$\\
$\taubub$ & $\sigma_\mathrm{LJ} (m/\epsilon_\mathrm{LJ})^{1/2}$ & $7.7\,10^6$& $4.48$\\
$\taubub/\tauliq$ & --- & $350$ 	& $20.4$  
\end{tabular}
\end{ruledtabular}
\end{table}

\section{Discussion\label{sec:discussion}}

Regarding applications to geosciences, the present study sheds light on the conditions at which bubble collapse is expected. In Ref.~\onlinecite{Marti_effect_2012}, Marti~{\etal} developed a model related to the present one to determine the temperature at which bubble collapse occur when heating a small cavity. They made the following comment: ``We know that $T_\mathrm{h,obs}$ [the observed temperature of bubble collapse] must lie between $T_\mathrm{bin}$ of the “bubble binodal” and $T_\mathrm{sp}$ of the “bubble spinodal”, but it cannot be predicted because of the inherent uncertainty associated
with a metastable state.'' For this reason, in the analysis of geophysical data on bubble collapse in minerals, in order to minimize the uncertainty, it is customary to assume that bubble collapse occurs at a temperature $(T_\mathrm{bin}+T_\mathrm{sp})/2$ (see for instance Ref.~\onlinecite{Guillerm_physical_2025}). From Fig.~\ref{fig:dens}, we know that for cavity volumes relevant to geosciences (i.e. $V>\qty{1}{\mu m^3}$), spontaneous bubble collapse requires $\dav\approx \ds$. Assuming as in experiment that the system is heated at constant cavity volume, this translates into a temperature of bubble disappearance very close to $T_\mathrm{s}$ (typically within less than \qty{0.1}{\degreeCelsius}). It would thus seem better to use $T_\mathrm{sp}$ rather than $(T_\mathrm{bin}+T_\mathrm{sp})/2$. Unfortunately, all our reasoning is based on the implicit assumption that nucleation occurs homogeneously. In real systems, impurities or wall defects may reduce the energy barrier for bubble collapse, which may then occur away from $T_\mathrm{sp}$ (whose value could also be modified). Without more information on the type of nucleation process, it is therefore better to use the conservative value $(T_\mathrm{bin}+T_\mathrm{sp})/2$.

The fact that the system will change state only very close to its limit of stability is reminiscent of the classical picture of capillary condensation~\cite{Cole_excitation_1974,Saam_excitations_1975}. When a cylindrical pore is exposed to a subsaturated vapor, a liquid film forms if the liquid wets the pore walls. The film thickness grows with increasing vapor pressure, and the film becomes metastable with respect to the fully filled pore. However, complete filling does not occur immediately, because it requires a macroscopic deformation of the film which is energetically costly. Filling occurs only close to an instability of the film which is reached for a critical thickness. In contrast, emptying may occur at equilibrium between the film and filled states. This is the basis for the explanation of adsorption hysteresis in small pores.

Regarding simulations, our theoretical approach provides a detailed picture for the conditions necessary to observe phase flipping. They are qualitatively confirmed by direct simulations of a Lennard-Jones system. This illustrates how finite size effects modify the observation of phase coexistence. A similar situation is observed in other systems. For instance, ST2 is a model of water that predicts a first-order liquid-liquid transition at low temperature, terminating at a liquid-liquid critical point (LLCP). NPT simulations of small systems near the LLCP found that ``the entire system flips rapidly between liquid states of high and low density''\cite{Kesselring_nanoscale_2012}. Note that, in that case, flipping occurred between states with only one phase, due to the NPT conditions. In our NVT case, flipping occurs between a state with one phase (fully liquid), and a state with two phases (liquid and bubble). 

Although the simulation results are in good qualitative agreement with the model predictions, there are large quantitative differences. The density at the bubble binodal, and the corresponding energy barrier, are significantly less than the model predictions. This results in vastly faster phase oscillations in the simulations compared to the model expectations. A similar observation was made by Menzl~{\etal} in their study of cavitation in water~\cite{Menzl_molecular_2016}: plain CNT underestimates the cavitation rates by more than 15 orders of magnitude. One of the reasons for this discrepancy is that CNT ignores the effect of curvature on surface tension. This becomes a poor approximation for small bubbles, when the thickness of the liquid-vapor interface cannot be neglected compared to the radius~\cite{Menzl_molecular_2016,Bruot_curvature_2016}. A decrease in surface tension due to curvature would be consistent with the smaller characteristic bubble volumes (Fig.~\ref{fig:Vcomp}) and lower barriers (Fig.~\ref{fig:LJcurves}) observed in simulations. Additionally, the linear approximation for the chemical potential (Eq.~\ref{eq:mu}) deteriorates at the low densities reached in the simulations~\cite{MacDowell_nucleation_2006}. This could be improved by using a specific equation of state, but would reduce generality. Another limitation is that, for the small energy barrier found in the present simulation ($\simeq 1.1\, \kB T$), Kramers' theory is expected to become inaccurate. It would be interesting to see how the comparison with the model evolves when the cavity volume $V$ is increased, which will make the barrier higher. This will however require longer simulations, and possibly advanced sampling techniques to reach configurations corresponding to the energy barrier. Future work could investigate this further, as well as the role played by the assumption of a spherical bubble and by the choice of reaction coordinate\cite{Puibasset_molecularsized_2025}.

\begin{acknowledgments}

We thank Luis G. MacDowell for bringing several references to our attention. F.C. dedicates this work to the memory of Arezki Boudaoud, for a long-standing friendship and stimulating discussions. C.V. acknowledges fundings IHRC22/00002 and Proyecto PID2022-140407NB-C21 funded by
MCIN/AEI /10.13039/501100011033 and FEDER, UE.
\end{acknowledgments}

\section*{Availability Statement}
The data that support the findings of this study are available from the corresponding author upon reasonable request.

\bibliography{references}

\begin{thebibliography}{51}%
\makeatletter
\providecommand \@ifxundefined [1]{%
 \@ifx{#1\undefined}
}%
\providecommand \@ifnum [1]{%
 \ifnum #1\expandafter \@firstoftwo
 \else \expandafter \@secondoftwo
 \fi
}%
\providecommand \@ifx [1]{%
 \ifx #1\expandafter \@firstoftwo
 \else \expandafter \@secondoftwo
 \fi
}%
\providecommand \natexlab [1]{#1}%
\providecommand \enquote  [1]{``#1''}%
\providecommand \bibnamefont  [1]{#1}%
\providecommand \bibfnamefont [1]{#1}%
\providecommand \citenamefont [1]{#1}%
\providecommand \href@noop [0]{\@secondoftwo}%
\providecommand \href [0]{\begingroup \@sanitize@url \@href}%
\providecommand \@href[1]{\@@startlink{#1}\@@href}%
\providecommand \@@href[1]{\endgroup#1\@@endlink}%
\providecommand \@sanitize@url [0]{\catcode `\\12\catcode `\$12\catcode
  `\&12\catcode `\#12\catcode `\^12\catcode `\_12\catcode `\%12\relax}%
\providecommand \@@startlink[1]{}%
\providecommand \@@endlink[0]{}%
\providecommand \url  [0]{\begingroup\@sanitize@url \@url }%
\providecommand \@url [1]{\endgroup\@href {#1}{\urlprefix }}%
\providecommand \urlprefix  [0]{URL }%
\providecommand \Eprint [0]{\href }%
\providecommand \doibase [0]{http://dx.doi.org/}%
\providecommand \selectlanguage [0]{\@gobble}%
\providecommand \bibinfo  [0]{\@secondoftwo}%
\providecommand \bibfield  [0]{\@secondoftwo}%
\providecommand \translation [1]{[#1]}%
\providecommand \BibitemOpen [0]{}%
\providecommand \bibitemStop [0]{}%
\providecommand \bibitemNoStop [0]{.\EOS\space}%
\providecommand \EOS [0]{\spacefactor3000\relax}%
\providecommand \BibitemShut  [1]{\csname bibitem#1\endcsname}%
\let\auto@bib@innerbib\@empty
\bibitem [{\citenamefont {Rollins}\ and\ \citenamefont
  {Dill}(2014)}]{Rollins_general_2014}%
  \BibitemOpen
  \bibfield  {author} {\bibinfo {author} {\bibfnamefont {G.~C.}\ \bibnamefont
  {Rollins}}\ and\ \bibinfo {author} {\bibfnamefont {K.~A.}\ \bibnamefont
  {Dill}},\ }\href {\doibase 10.1021/ja5049434} {\bibfield  {journal} {\bibinfo
   {journal} {J. Am. Chem. Soc.}\ }\textbf {\bibinfo {volume} {136}},\ \bibinfo
  {pages} {11420} (\bibinfo {year} {2014})}\BibitemShut {NoStop}%
\bibitem [{\citenamefont {Lyons}\ \emph {et~al.}(2024)\citenamefont {Lyons},
  \citenamefont {Devi}, \citenamefont {Hoffer},\ and\ \citenamefont
  {Woodside}}]{Lyons_quantifying_2024}%
  \BibitemOpen
  \bibfield  {author} {\bibinfo {author} {\bibfnamefont {A.}~\bibnamefont
  {Lyons}}, \bibinfo {author} {\bibfnamefont {A.}~\bibnamefont {Devi}},
  \bibinfo {author} {\bibfnamefont {N.~Q.}\ \bibnamefont {Hoffer}}, \ and\
  \bibinfo {author} {\bibfnamefont {M.~T.}\ \bibnamefont {Woodside}},\ }\href
  {\doibase 10.1103/PhysRevX.14.011017} {\bibfield  {journal} {\bibinfo
  {journal} {Phys. Rev. X}\ }\textbf {\bibinfo {volume} {14}},\ \bibinfo
  {pages} {011017} (\bibinfo {year} {2024})}\BibitemShut {NoStop}%
\bibitem [{\citenamefont {McCann}\ \emph {et~al.}(1999)\citenamefont {McCann},
  \citenamefont {Dykman},\ and\ \citenamefont
  {Golding}}]{McCann_thermally_1999}%
  \BibitemOpen
  \bibfield  {author} {\bibinfo {author} {\bibfnamefont {L.~I.}\ \bibnamefont
  {McCann}}, \bibinfo {author} {\bibfnamefont {M.}~\bibnamefont {Dykman}}, \
  and\ \bibinfo {author} {\bibfnamefont {B.}~\bibnamefont {Golding}},\ }\href
  {\doibase 10.1038/45492} {\bibfield  {journal} {\bibinfo  {journal} {Nature}\
  }\textbf {\bibinfo {volume} {402}},\ \bibinfo {pages} {785} (\bibinfo {year}
  {1999})}\BibitemShut {NoStop}%
\bibitem [{\citenamefont {Chupeau}\ \emph {et~al.}(2020)\citenamefont
  {Chupeau}, \citenamefont {Gladrow}, \citenamefont {Chepelianskii},
  \citenamefont {Keyser},\ and\ \citenamefont
  {Trizac}}]{Chupeau_optimizing_2020}%
  \BibitemOpen
  \bibfield  {author} {\bibinfo {author} {\bibfnamefont {M.}~\bibnamefont
  {Chupeau}}, \bibinfo {author} {\bibfnamefont {J.}~\bibnamefont {Gladrow}},
  \bibinfo {author} {\bibfnamefont {A.}~\bibnamefont {Chepelianskii}}, \bibinfo
  {author} {\bibfnamefont {U.~F.}\ \bibnamefont {Keyser}}, \ and\ \bibinfo
  {author} {\bibfnamefont {E.}~\bibnamefont {Trizac}},\ }\href {\doibase
  10.1073/pnas.1910677116} {\bibfield  {journal} {\bibinfo  {journal} {Proc.
  Natl. Acad. Sci. U.S.A.}\ }\textbf {\bibinfo {volume} {117}},\ \bibinfo
  {pages} {1383} (\bibinfo {year} {2020})}\BibitemShut {NoStop}%
\bibitem [{\citenamefont {Debenedetti}(1996)}]{Debenedetti_metastable_1996}%
  \BibitemOpen
  \bibfield  {author} {\bibinfo {author} {\bibfnamefont {P.~G.}\ \bibnamefont
  {Debenedetti}},\ }\href@noop {} {\emph {\bibinfo {title} {Metastable
  Liquids}}}\ (\bibinfo  {publisher} {Princeton University Press},\ \bibinfo
  {year} {1996})\BibitemShut {NoStop}%
\bibitem [{\citenamefont {Caupin}\ and\ \citenamefont
  {Grisenti}(2026)}]{Caupin_nucleation_2026}%
  \BibitemOpen
  \bibfield  {author} {\bibinfo {author} {\bibfnamefont {F.}~\bibnamefont
  {Caupin}}\ and\ \bibinfo {author} {\bibfnamefont {R.~E.}\ \bibnamefont
  {Grisenti}},\ }\href {\doibase 10.1021/acs.jpclett.5c03082} {\bibfield
  {journal} {\bibinfo  {journal} {J. Phys. Chem. Lett.}\ } (\bibinfo {year}
  {2026}),\ 10.1021/acs.jpclett.5c03082}\BibitemShut {NoStop}%
\bibitem [{\citenamefont {Gibbs}(1878)}]{Gibbs_equilibrium_1878}%
  \BibitemOpen
  \bibfield  {author} {\bibinfo {author} {\bibfnamefont {J.~W.}\ \bibnamefont
  {Gibbs}},\ }\href@noop {} {\bibfield  {journal} {\bibinfo  {journal} {Trans.
  Conn. Acad.}\ }\textbf {\bibinfo {volume} {3}},\ \bibinfo {pages} {343}
  (\bibinfo {year} {1878})}\BibitemShut {NoStop}%
\bibitem [{\citenamefont {Volmer}\ and\ \citenamefont
  {Weber}(1926)}]{Volmer_Keimbildung_1926}%
  \BibitemOpen
  \bibfield  {author} {\bibinfo {author} {\bibfnamefont {M.}~\bibnamefont
  {Volmer}}\ and\ \bibinfo {author} {\bibfnamefont {A.}~\bibnamefont {Weber}},\
  }\href {\doibase 10.1515/zpch-1926-11927} {\bibfield  {journal} {\bibinfo
  {journal} {Zeitschrift f\"ur Physikalische Chemie}\ }\textbf {\bibinfo
  {volume} {119U}},\ \bibinfo {pages} {277} (\bibinfo {year}
  {1926})}\BibitemShut {NoStop}%
\bibitem [{\citenamefont {Fisher}(1948)}]{Fisher_fracture_1948}%
  \BibitemOpen
  \bibfield  {author} {\bibinfo {author} {\bibfnamefont {J.~C.}\ \bibnamefont
  {Fisher}},\ }\href {\doibase 10.1063/1.1698012} {\bibfield  {journal}
  {\bibinfo  {journal} {J. Appl. Phys.}\ }\textbf {\bibinfo {volume} {19}},\
  \bibinfo {pages} {1062} (\bibinfo {year} {Novembre 00, 1948})}\BibitemShut
  {NoStop}%
\bibitem [{\citenamefont {Zheng}\ \emph {et~al.}(1991)\citenamefont {Zheng},
  \citenamefont {Durben}, \citenamefont {Wolf},\ and\ \citenamefont
  {Angell}}]{Zheng_liquids_1991}%
  \BibitemOpen
  \bibfield  {author} {\bibinfo {author} {\bibfnamefont {Q.}~\bibnamefont
  {Zheng}}, \bibinfo {author} {\bibfnamefont {D.~J.}\ \bibnamefont {Durben}},
  \bibinfo {author} {\bibfnamefont {G.~H.}\ \bibnamefont {Wolf}}, \ and\
  \bibinfo {author} {\bibfnamefont {C.~A.}\ \bibnamefont {Angell}},\ }\href
  {\doibase 10.1126/science.254.5033.829} {\bibfield  {journal} {\bibinfo
  {journal} {Science}\ }\textbf {\bibinfo {volume} {254}},\ \bibinfo {pages}
  {829} (\bibinfo {year} {1991})}\BibitemShut {NoStop}%
\bibitem [{\citenamefont {Azouzi}\ \emph {et~al.}(2012)\citenamefont {Azouzi},
  \citenamefont {Ramboz}, \citenamefont {Lenain},\ and\ \citenamefont
  {Caupin}}]{Azouzi_coherent_2012}%
  \BibitemOpen
  \bibfield  {author} {\bibinfo {author} {\bibfnamefont {M.~E.~M.}\
  \bibnamefont {Azouzi}}, \bibinfo {author} {\bibfnamefont {C.}~\bibnamefont
  {Ramboz}}, \bibinfo {author} {\bibfnamefont {J.-F.}\ \bibnamefont {Lenain}},
  \ and\ \bibinfo {author} {\bibfnamefont {F.}~\bibnamefont {Caupin}},\ }\href
  {\doibase 10.1038/nphys2475} {\bibfield  {journal} {\bibinfo  {journal}
  {Nature Physics}\ }\textbf {\bibinfo {volume} {9}},\ \bibinfo {pages} {38}
  (\bibinfo {year} {2012})}\BibitemShut {NoStop}%
\bibitem [{\citenamefont {Menzl}\ \emph {et~al.}(2016)\citenamefont {Menzl},
  \citenamefont {Gonzalez}, \citenamefont {Geiger}, \citenamefont {Caupin},
  \citenamefont {Abascal}, \citenamefont {Valeriani},\ and\ \citenamefont
  {Dellago}}]{Menzl_molecular_2016}%
  \BibitemOpen
  \bibfield  {author} {\bibinfo {author} {\bibfnamefont {G.}~\bibnamefont
  {Menzl}}, \bibinfo {author} {\bibfnamefont {M.~A.}\ \bibnamefont {Gonzalez}},
  \bibinfo {author} {\bibfnamefont {P.}~\bibnamefont {Geiger}}, \bibinfo
  {author} {\bibfnamefont {F.}~\bibnamefont {Caupin}}, \bibinfo {author}
  {\bibfnamefont {J.~L.~F.}\ \bibnamefont {Abascal}}, \bibinfo {author}
  {\bibfnamefont {C.}~\bibnamefont {Valeriani}}, \ and\ \bibinfo {author}
  {\bibfnamefont {C.}~\bibnamefont {Dellago}},\ }\href {\doibase
  10.1073/pnas.1608421113} {\bibfield  {journal} {\bibinfo  {journal}
  {Proceedings of the National Academy of Sciences}\ }\textbf {\bibinfo
  {volume} {113}},\ \bibinfo {pages} {13582} (\bibinfo {year}
  {2016})}\BibitemShut {NoStop}%
\bibitem [{\citenamefont {Hurai}\ \emph {et~al.}(2015)\citenamefont {Hurai},
  \citenamefont {Huraiov{\'a}}, \citenamefont {Slobodn{\'i}k},\ and\
  \citenamefont {Thomas}}]{Hurai_geofluids_2015}%
  \BibitemOpen
  \bibinfo {editor} {\bibfnamefont {V.}~\bibnamefont {Hurai}}, \bibinfo
  {editor} {\bibfnamefont {M.}~\bibnamefont {Huraiov{\'a}}}, \bibinfo {editor}
  {\bibfnamefont {M.}~\bibnamefont {Slobodn{\'i}k}}, \ and\ \bibinfo {editor}
  {\bibfnamefont {R.}~\bibnamefont {Thomas}},\ eds.,\ \href {\doibase
  10.1016/C2014-0-03099-7} {\emph {\bibinfo {title} {Geofluids}}},\
  Vapor-{{Liquid Equilibrium Data Bibliography}}\ (\bibinfo  {publisher}
  {Elsevier},\ \bibinfo {year} {2015})\BibitemShut {NoStop}%
\bibitem [{\citenamefont {Roedder}(1984)}]{Roedder_fluid_1984}%
  \BibitemOpen
  \bibfield  {author} {\bibinfo {author} {\bibfnamefont {E.}~\bibnamefont
  {Roedder}},\ }\href {\doibase 10.1515/9781501508271} {\emph {\bibinfo {title}
  {Fluid Inclusions}}},\ \bibinfo {series} {Reviews in {{Mineralogy}}}\
  No.~\bibinfo {number} {12}\ (\bibinfo  {publisher} {De Gruyter},\ \bibinfo
  {year} {1984})\BibitemShut {NoStop}%
\bibitem [{\citenamefont {Kr{\"u}ger}\ \emph {et~al.}(2011)\citenamefont
  {Kr{\"u}ger}, \citenamefont {Marti}, \citenamefont {Staub}, \citenamefont
  {Fleitmann},\ and\ \citenamefont {Frenz}}]{Kruger_liquid_2011}%
  \BibitemOpen
  \bibfield  {author} {\bibinfo {author} {\bibfnamefont {Y.}~\bibnamefont
  {Kr{\"u}ger}}, \bibinfo {author} {\bibfnamefont {D.}~\bibnamefont {Marti}},
  \bibinfo {author} {\bibfnamefont {R.~H.}\ \bibnamefont {Staub}}, \bibinfo
  {author} {\bibfnamefont {D.}~\bibnamefont {Fleitmann}}, \ and\ \bibinfo
  {author} {\bibfnamefont {M.}~\bibnamefont {Frenz}},\ }\href {\doibase
  10.1016/j.chemgeo.2011.07.009} {\bibfield  {journal} {\bibinfo  {journal}
  {Chemical Geology}\ }\textbf {\bibinfo {volume} {289}},\ \bibinfo {pages}
  {39} (\bibinfo {year} {2011})}\BibitemShut {NoStop}%
\bibitem [{\citenamefont {Guillerm}\ \emph {et~al.}(2025)\citenamefont
  {Guillerm}, \citenamefont {Lowenstein}, \citenamefont {Gardien},
  \citenamefont {Brauer}, \citenamefont {Kr{\"u}ger}, \citenamefont {Arnuk},\
  and\ \citenamefont {Caupin}}]{Guillerm_physical_2025}%
  \BibitemOpen
  \bibfield  {author} {\bibinfo {author} {\bibfnamefont {E.}~\bibnamefont
  {Guillerm}}, \bibinfo {author} {\bibfnamefont {T.~K.}\ \bibnamefont
  {Lowenstein}}, \bibinfo {author} {\bibfnamefont {V.}~\bibnamefont {Gardien}},
  \bibinfo {author} {\bibfnamefont {A.}~\bibnamefont {Brauer}}, \bibinfo
  {author} {\bibfnamefont {Y.}~\bibnamefont {Kr{\"u}ger}}, \bibinfo {author}
  {\bibfnamefont {W.~D.}\ \bibnamefont {Arnuk}}, \ and\ \bibinfo {author}
  {\bibfnamefont {F.}~\bibnamefont {Caupin}},\ }\href {\doibase
  10.2475/001c.130836} {\bibfield  {journal} {\bibinfo  {journal} {American
  Journal of Science}\ }\textbf {\bibinfo {volume} {325}} (\bibinfo {year}
  {2025}),\ 10.2475/001c.130836}\BibitemShut {NoStop}%
\bibitem [{\citenamefont {Marti}\ \emph {et~al.}(2012)\citenamefont {Marti},
  \citenamefont {Kr{\"u}ger}, \citenamefont {Fleitmann}, \citenamefont
  {Frenz},\ and\ \citenamefont {Ri{\v c}ka}}]{Marti_effect_2012}%
  \BibitemOpen
  \bibfield  {author} {\bibinfo {author} {\bibfnamefont {D.}~\bibnamefont
  {Marti}}, \bibinfo {author} {\bibfnamefont {Y.}~\bibnamefont {Kr{\"u}ger}},
  \bibinfo {author} {\bibfnamefont {D.}~\bibnamefont {Fleitmann}}, \bibinfo
  {author} {\bibfnamefont {M.}~\bibnamefont {Frenz}}, \ and\ \bibinfo {author}
  {\bibfnamefont {J.}~\bibnamefont {Ri{\v c}ka}},\ }\href {\doibase
  10.1016/j.fluid.2011.08.010} {\bibfield  {journal} {\bibinfo  {journal}
  {Fluid Phase Equilib.}\ }\textbf {\bibinfo {volume} {314}},\ \bibinfo {pages}
  {13} (\bibinfo {year} {2012})}\BibitemShut {NoStop}%
\bibitem [{\citenamefont {Wilhelmsen}\ \emph {et~al.}(2014)\citenamefont
  {Wilhelmsen}, \citenamefont {Bedeaux}, \citenamefont {Kjelstrup},\ and\
  \citenamefont {Reguera}}]{Wilhelmsen_communication_2014}%
  \BibitemOpen
  \bibfield  {author} {\bibinfo {author} {\bibfnamefont {{\O}.}~\bibnamefont
  {Wilhelmsen}}, \bibinfo {author} {\bibfnamefont {D.}~\bibnamefont {Bedeaux}},
  \bibinfo {author} {\bibfnamefont {S.}~\bibnamefont {Kjelstrup}}, \ and\
  \bibinfo {author} {\bibfnamefont {D.}~\bibnamefont {Reguera}},\ }\href
  {\doibase 10.1063/1.4893701} {\bibfield  {journal} {\bibinfo  {journal} {The
  Journal of Chemical Physics}\ }\textbf {\bibinfo {volume} {141}},\ \bibinfo
  {pages} {071103} (\bibinfo {year} {2014})}\BibitemShut {NoStop}%
\bibitem [{\citenamefont {Wilhelmsen}\ and\ \citenamefont
  {Reguera}(2015)}]{Wilhelmsen_evaluation_2015}%
  \BibitemOpen
  \bibfield  {author} {\bibinfo {author} {\bibfnamefont {{\O}.}~\bibnamefont
  {Wilhelmsen}}\ and\ \bibinfo {author} {\bibfnamefont {D.}~\bibnamefont
  {Reguera}},\ }\href {\doibase 10.1063/1.4907367} {\bibfield  {journal}
  {\bibinfo  {journal} {The Journal of Chemical Physics}\ }\textbf {\bibinfo
  {volume} {142}},\ \bibinfo {pages} {064703} (\bibinfo {year}
  {2015})}\BibitemShut {NoStop}%
\bibitem [{\citenamefont {Vincent}\ and\ \citenamefont
  {Marmottant}(2017)}]{Vincent_statics_2017}%
  \BibitemOpen
  \bibfield  {author} {\bibinfo {author} {\bibfnamefont {O.}~\bibnamefont
  {Vincent}}\ and\ \bibinfo {author} {\bibfnamefont {P.}~\bibnamefont
  {Marmottant}},\ }\href {\doibase 10.1017/jfm.2017.487} {\bibfield  {journal}
  {\bibinfo  {journal} {J. Fluid Mech.}\ }\textbf {\bibinfo {volume} {827}},\
  \bibinfo {pages} {194} (\bibinfo {year} {2017})}\BibitemShut {NoStop}%
\bibitem [{\citenamefont {Caupin}(2022)}]{Caupin_effects_2022}%
  \BibitemOpen
  \bibfield  {author} {\bibinfo {author} {\bibfnamefont {F.}~\bibnamefont
  {Caupin}},\ }\href {\doibase 10.1063/5.0098969} {\bibfield  {journal}
  {\bibinfo  {journal} {J. Chem. Phys.}\ }\textbf {\bibinfo {volume} {157}},\
  \bibinfo {pages} {054506} (\bibinfo {year} {2022})}\BibitemShut {NoStop}%
\bibitem [{\citenamefont {MacDowell}\ \emph {et~al.}(2006)\citenamefont
  {MacDowell}, \citenamefont {Shen},\ and\ \citenamefont
  {Errington}}]{MacDowell_nucleation_2006}%
  \BibitemOpen
  \bibfield  {author} {\bibinfo {author} {\bibfnamefont {L.~G.}\ \bibnamefont
  {MacDowell}}, \bibinfo {author} {\bibfnamefont {V.~K.}\ \bibnamefont {Shen}},
  \ and\ \bibinfo {author} {\bibfnamefont {J.~R.}\ \bibnamefont {Errington}},\
  }\href {\doibase 10.1063/1.2218845} {\bibfield  {journal} {\bibinfo
  {journal} {The Journal of Chemical Physics}\ }\textbf {\bibinfo {volume}
  {125}},\ \bibinfo {pages} {034705} (\bibinfo {year} {2006})}\BibitemShut
  {NoStop}%
\bibitem [{\citenamefont {Glavatskiy}\ \emph {et~al.}(2013)\citenamefont
  {Glavatskiy}, \citenamefont {Reguera},\ and\ \citenamefont
  {Bedeaux}}]{Glavatskiy_effect_2013}%
  \BibitemOpen
  \bibfield  {author} {\bibinfo {author} {\bibfnamefont {K.~S.}\ \bibnamefont
  {Glavatskiy}}, \bibinfo {author} {\bibfnamefont {D.}~\bibnamefont {Reguera}},
  \ and\ \bibinfo {author} {\bibfnamefont {D.}~\bibnamefont {Bedeaux}},\ }\href
  {\doibase 10.1063/1.4807323} {\bibfield  {journal} {\bibinfo  {journal} {The
  Journal of Chemical Physics}\ }\textbf {\bibinfo {volume} {138}},\ \bibinfo
  {pages} {204708} (\bibinfo {year} {2013})}\BibitemShut {NoStop}%
\bibitem [{\citenamefont {{Llamas-Jaramillo}}\ \emph
  {et~al.}(2026)\citenamefont {{Llamas-Jaramillo}}, \citenamefont {Latella},\
  and\ \citenamefont {Reguera}}]{Llamas-Jaramillo_freeenergy_2026}%
  \BibitemOpen
  \bibfield  {author} {\bibinfo {author} {\bibfnamefont {A.}~\bibnamefont
  {{Llamas-Jaramillo}}}, \bibinfo {author} {\bibfnamefont {I.}~\bibnamefont
  {Latella}}, \ and\ \bibinfo {author} {\bibfnamefont {D.}~\bibnamefont
  {Reguera}},\ }\href {\doibase 10.1063/5.0309755} {\bibfield  {journal}
  {\bibinfo  {journal} {The Journal of Chemical Physics}\ }\textbf {\bibinfo
  {volume} {164}},\ \bibinfo {pages} {034102} (\bibinfo {year}
  {2026})}\BibitemShut {NoStop}%
\bibitem [{\citenamefont {{The International Association for the Properties of
  Water and
  Steam}}(1992)}]{TheInternationalAssociationforthePropertiesofWaterandSteam_revised_1992}%
  \BibitemOpen
  \bibfield  {author} {\bibinfo {author} {\bibnamefont {{The International
  Association for the Properties of Water and Steam}}},\ }\href@noop {} {\emph
  {\bibinfo {title} {Revised supplementary release on saturation properties of
  ordinary water substance}}},\ \bibinfo {type} {Tech. Rep.}\ \bibinfo {number}
  {IAPWS SR1-86(1992)}\ (\bibinfo  {institution} {The International Association
  for the Properties of Water and Steam},\ \bibinfo {year} {1992})\BibitemShut
  {NoStop}%
\bibitem [{\citenamefont {{The International Association for the Properties of
  Water and
  Steam}}(2014)}]{TheInternationalAssociationforthePropertiesofWaterandSteam_revised_2014}%
  \BibitemOpen
  \bibfield  {author} {\bibinfo {author} {\bibnamefont {{The International
  Association for the Properties of Water and Steam}}},\ }\href@noop {} {\emph
  {\bibinfo {title} {Revised release on surface tension of ordinary water
  substance}}},\ \bibinfo {type} {Tech. Rep.}\ \bibinfo {number} {R1-76(2014)}\
  (\bibinfo  {institution} {{The International Association for the Properties
  of Water and Steam}},\ \bibinfo {address} {Moscow},\ \bibinfo {year}
  {2014})\BibitemShut {NoStop}%
\bibitem [{\citenamefont {{The International Association for the Properties of
  Water and
  Steam}}(2018)}]{TheInternationalAssociationforthePropertiesofWaterandSteam_revised_2018}%
  \BibitemOpen
  \bibfield  {author} {\bibinfo {author} {\bibnamefont {{The International
  Association for the Properties of Water and Steam}}},\ }\href@noop {} {\emph
  {\bibinfo {title} {Revised release on the {IAPWS} formulation 1995 for the
  thermodynamic properties of ordinary water substance for general and
  scientific use}}},\ \bibinfo {type} {Tech. Rep.}\ \bibinfo {number} {IAPWS
  R6-95(2018)}\ (\bibinfo  {institution} {The International Association for the
  Properties of Water and Steam},\ \bibinfo {year} {2018})\BibitemShut
  {NoStop}%
\bibitem [{\citenamefont {Kramers}(1940)}]{Kramers_Brownian_1940}%
  \BibitemOpen
  \bibfield  {author} {\bibinfo {author} {\bibfnamefont {H.~A.}\ \bibnamefont
  {Kramers}},\ }\href {\doibase 10.1016/S0031-8914(40)90098-2} {\bibfield
  {journal} {\bibinfo  {journal} {Physica}\ }\textbf {\bibinfo {volume} {7}},\
  \bibinfo {pages} {284} (\bibinfo {year} {1940})}\BibitemShut {NoStop}%
\bibitem [{\citenamefont {Schulten}\ \emph {et~al.}(1981)\citenamefont
  {Schulten}, \citenamefont {Schulten},\ and\ \citenamefont
  {Szabo}}]{Schulten_dynamics_1981}%
  \BibitemOpen
  \bibfield  {author} {\bibinfo {author} {\bibfnamefont {K.}~\bibnamefont
  {Schulten}}, \bibinfo {author} {\bibfnamefont {Z.}~\bibnamefont {Schulten}},
  \ and\ \bibinfo {author} {\bibfnamefont {A.}~\bibnamefont {Szabo}},\ }\href
  {\doibase 10.1063/1.441684} {\bibfield  {journal} {\bibinfo  {journal} {J.
  Chem. Phys.}\ }\textbf {\bibinfo {volume} {74}},\ \bibinfo {pages} {4426}
  (\bibinfo {year} {1981})}\BibitemShut {NoStop}%
\bibitem [{\citenamefont {Brennen}(1995)}]{Brennen_cavitation_1995}%
  \BibitemOpen
  \bibfield  {author} {\bibinfo {author} {\bibfnamefont {C.~E.}\ \bibnamefont
  {Brennen}},\ }\href {\doibase 10.1093/oso/9780195094091.001.0001} {\emph
  {\bibinfo {title} {Cavitation {{And Bubble Dynamics}}}}}\ (\bibinfo
  {publisher} {Oxford University Press},\ \bibinfo {year} {1995})\BibitemShut
  {NoStop}%
\bibitem [{\citenamefont {Park}\ \emph {et~al.}(2001)\citenamefont {Park},
  \citenamefont {Weng},\ and\ \citenamefont {Tien}}]{Park_molecular_2001}%
  \BibitemOpen
  \bibfield  {author} {\bibinfo {author} {\bibfnamefont {S.}~\bibnamefont
  {Park}}, \bibinfo {author} {\bibfnamefont {J.}~\bibnamefont {Weng}}, \ and\
  \bibinfo {author} {\bibfnamefont {C.}~\bibnamefont {Tien}},\ }\href {\doibase
  10.1016/S0017-9310(00)00244-1} {\bibfield  {journal} {\bibinfo  {journal}
  {International Journal of Heat and Mass Transfer}\ }\textbf {\bibinfo
  {volume} {44}},\ \bibinfo {pages} {1849} (\bibinfo {year}
  {2001})}\BibitemShut {NoStop}%
\bibitem [{\citenamefont {Neimark}\ and\ \citenamefont
  {Vishnyakov}(2005)}]{Neimark_birth_2005}%
  \BibitemOpen
  \bibfield  {author} {\bibinfo {author} {\bibfnamefont {A.~V.}\ \bibnamefont
  {Neimark}}\ and\ \bibinfo {author} {\bibfnamefont {A.}~\bibnamefont
  {Vishnyakov}},\ }\href {\doibase 10.1063/1.1829040} {\bibfield  {journal}
  {\bibinfo  {journal} {The Journal of Chemical Physics}\ }\textbf {\bibinfo
  {volume} {122}},\ \bibinfo {pages} {054707} (\bibinfo {year}
  {2005})}\BibitemShut {NoStop}%
\bibitem [{\citenamefont {Neimark}\ and\ \citenamefont
  {Vishnyakov}(2006)}]{Neimark_phase_2006}%
  \BibitemOpen
  \bibfield  {author} {\bibinfo {author} {\bibfnamefont {A.~V.}\ \bibnamefont
  {Neimark}}\ and\ \bibinfo {author} {\bibfnamefont {A.}~\bibnamefont
  {Vishnyakov}},\ }\href {\doibase 10.1021/jp056407d} {\bibfield  {journal}
  {\bibinfo  {journal} {J. Phys. Chem. B}\ }\textbf {\bibinfo {volume} {110}},\
  \bibinfo {pages} {9403} (\bibinfo {year} {2006})}\BibitemShut {NoStop}%
\bibitem [{\citenamefont {Chen}\ \emph {et~al.}(2001)\citenamefont {Chen},
  \citenamefont {Siepmann},\ and\ \citenamefont {Klein}}]{Chen2001}%
  \BibitemOpen
  \bibfield  {author} {\bibinfo {author} {\bibfnamefont {B.}~\bibnamefont
  {Chen}}, \bibinfo {author} {\bibfnamefont {J.~I.}\ \bibnamefont {Siepmann}},
  \ and\ \bibinfo {author} {\bibfnamefont {M.~L.}\ \bibnamefont {Klein}},\
  }\href {\doibase 10.1021/jp011950p} {\bibfield  {journal} {\bibinfo
  {journal} {The Journal of Physical Chemistry B}\ }\textbf {\bibinfo {volume}
  {105}},\ \bibinfo {pages} {9840} (\bibinfo {year} {2001})}\BibitemShut
  {NoStop}%
\bibitem [{\citenamefont {Stephan}\ \emph {et~al.}(2019)\citenamefont
  {Stephan}, \citenamefont {Thol}, \citenamefont {Vrabec},\ and\ \citenamefont
  {Hasse}}]{Stephan2019}%
  \BibitemOpen
  \bibfield  {author} {\bibinfo {author} {\bibfnamefont {S.}~\bibnamefont
  {Stephan}}, \bibinfo {author} {\bibfnamefont {M.}~\bibnamefont {Thol}},
  \bibinfo {author} {\bibfnamefont {J.}~\bibnamefont {Vrabec}}, \ and\ \bibinfo
  {author} {\bibfnamefont {H.}~\bibnamefont {Hasse}},\ }\href {\doibase
  10.1021/acs.jcim.9b00620} {\bibfield  {journal} {\bibinfo  {journal} {Journal
  of Chemical Information and Modeling}\ }\textbf {\bibinfo {volume} {59}},\
  \bibinfo {pages} {4248} (\bibinfo {year} {2019})}\BibitemShut {NoStop}%
\bibitem [{\citenamefont {Lotfi}\ \emph {et~al.}(1992)\citenamefont {Lotfi},
  \citenamefont {Vrabec},\ and\ \citenamefont {Fischer}}]{Lotfi1992}%
  \BibitemOpen
  \bibfield  {author} {\bibinfo {author} {\bibfnamefont {A.}~\bibnamefont
  {Lotfi}}, \bibinfo {author} {\bibfnamefont {J.}~\bibnamefont {Vrabec}}, \
  and\ \bibinfo {author} {\bibfnamefont {J.}~\bibnamefont {Fischer}},\ }\href
  {\doibase 10.1080/00268979200101611} {\bibfield  {journal} {\bibinfo
  {journal} {Molecular Physics}\ }\textbf {\bibinfo {volume} {76}},\ \bibinfo
  {pages} {1319} (\bibinfo {year} {1992})}\BibitemShut {NoStop}%
\bibitem [{\citenamefont {Meier}\ \emph {et~al.}(2004)\citenamefont {Meier},
  \citenamefont {Laesecke},\ and\ \citenamefont
  {Kabelac}}]{Meier_transport_2004}%
  \BibitemOpen
  \bibfield  {author} {\bibinfo {author} {\bibfnamefont {K.}~\bibnamefont
  {Meier}}, \bibinfo {author} {\bibfnamefont {A.}~\bibnamefont {Laesecke}}, \
  and\ \bibinfo {author} {\bibfnamefont {S.}~\bibnamefont {Kabelac}},\ }\href
  {\doibase 10.1063/1.1770695} {\bibfield  {journal} {\bibinfo  {journal} {The
  Journal of Chemical Physics}\ }\textbf {\bibinfo {volume} {121}},\ \bibinfo
  {pages} {3671} (\bibinfo {year} {2004})}\BibitemShut {NoStop}%
\bibitem [{\citenamefont {Thompson}\ \emph {et~al.}(2022)\citenamefont
  {Thompson}, \citenamefont {Aktulga}, \citenamefont {Berger}, \citenamefont
  {Bolintineanu}, \citenamefont {Brown}, \citenamefont {Crozier}, \citenamefont
  {{in 't Veld}}, \citenamefont {Kohlmeyer}, \citenamefont {Moore},
  \citenamefont {Nguyen}, \citenamefont {Shan}, \citenamefont {Stevens},
  \citenamefont {Tranchida}, \citenamefont {Trott},\ and\ \citenamefont
  {Plimpton}}]{Thompson_lammps_2022}%
  \BibitemOpen
  \bibfield  {author} {\bibinfo {author} {\bibfnamefont {A.~P.}\ \bibnamefont
  {Thompson}}, \bibinfo {author} {\bibfnamefont {H.~M.}\ \bibnamefont
  {Aktulga}}, \bibinfo {author} {\bibfnamefont {R.}~\bibnamefont {Berger}},
  \bibinfo {author} {\bibfnamefont {D.~S.}\ \bibnamefont {Bolintineanu}},
  \bibinfo {author} {\bibfnamefont {W.~M.}\ \bibnamefont {Brown}}, \bibinfo
  {author} {\bibfnamefont {P.~S.}\ \bibnamefont {Crozier}}, \bibinfo {author}
  {\bibfnamefont {P.~J.}\ \bibnamefont {{in 't Veld}}}, \bibinfo {author}
  {\bibfnamefont {A.}~\bibnamefont {Kohlmeyer}}, \bibinfo {author}
  {\bibfnamefont {S.~G.}\ \bibnamefont {Moore}}, \bibinfo {author}
  {\bibfnamefont {T.~D.}\ \bibnamefont {Nguyen}}, \bibinfo {author}
  {\bibfnamefont {R.}~\bibnamefont {Shan}}, \bibinfo {author} {\bibfnamefont
  {M.~J.}\ \bibnamefont {Stevens}}, \bibinfo {author} {\bibfnamefont
  {J.}~\bibnamefont {Tranchida}}, \bibinfo {author} {\bibfnamefont
  {C.}~\bibnamefont {Trott}}, \ and\ \bibinfo {author} {\bibfnamefont {S.~J.}\
  \bibnamefont {Plimpton}},\ }\href {\doibase 10.1016/j.cpc.2021.108171}
  {\bibfield  {journal} {\bibinfo  {journal} {Computer Physics Communications}\
  }\textbf {\bibinfo {volume} {271}},\ \bibinfo {pages} {108171} (\bibinfo
  {year} {2022})}\BibitemShut {NoStop}%
\bibitem [{\citenamefont {Nosé}(1984)}]{Nose1984}%
  \BibitemOpen
  \bibfield  {author} {\bibinfo {author} {\bibfnamefont {S.}~\bibnamefont
  {Nosé}},\ }\href {\doibase 10.1063/1.447334} {\bibfield  {journal} {\bibinfo
   {journal} {Journal of Chemical Physics}\ }\textbf {\bibinfo {volume} {81}},\
  \bibinfo {pages} {511} (\bibinfo {year} {1984})}\BibitemShut {NoStop}%
\bibitem [{\citenamefont {Hoover}(1985)}]{Hoover1985}%
  \BibitemOpen
  \bibfield  {author} {\bibinfo {author} {\bibfnamefont {W.~G.}\ \bibnamefont
  {Hoover}},\ }\href {\doibase 10.1103/PhysRevA.31.1695} {\bibfield  {journal}
  {\bibinfo  {journal} {Physical Review A}\ }\textbf {\bibinfo {volume} {31}},\
  \bibinfo {pages} {1695} (\bibinfo {year} {1985})}\BibitemShut {NoStop}%
\bibitem [{\citenamefont {Martyna}\ \emph {et~al.}(1992)\citenamefont
  {Martyna}, \citenamefont {Klein},\ and\ \citenamefont
  {Tuckerman}}]{Martyna1992}%
  \BibitemOpen
  \bibfield  {author} {\bibinfo {author} {\bibfnamefont {G.~J.}\ \bibnamefont
  {Martyna}}, \bibinfo {author} {\bibfnamefont {M.~L.}\ \bibnamefont {Klein}},
  \ and\ \bibinfo {author} {\bibfnamefont {M.}~\bibnamefont {Tuckerman}},\
  }\href {\doibase 10.1063/1.463940} {\bibfield  {journal} {\bibinfo  {journal}
  {Journal of Chemical Physics}\ }\textbf {\bibinfo {volume} {97}},\ \bibinfo
  {pages} {2635} (\bibinfo {year} {1992})}\BibitemShut {NoStop}%
\bibitem [{\citenamefont {Wang}\ \emph {et~al.}(2009)\citenamefont {Wang},
  \citenamefont {Valeriani},\ and\ \citenamefont
  {Frenkel}}]{Wang_homogeneous_2009}%
  \BibitemOpen
  \bibfield  {author} {\bibinfo {author} {\bibfnamefont {Z.-J.}\ \bibnamefont
  {Wang}}, \bibinfo {author} {\bibfnamefont {C.}~\bibnamefont {Valeriani}}, \
  and\ \bibinfo {author} {\bibfnamefont {D.}~\bibnamefont {Frenkel}},\ }\href
  {\doibase 10.1021/jp807727p} {\bibfield  {journal} {\bibinfo  {journal} {J.
  Phys. Chem. B}\ }\textbf {\bibinfo {volume} {113}},\ \bibinfo {pages} {3776}
  (\bibinfo {year} {2009})}\BibitemShut {NoStop}%
\bibitem [{\citenamefont {Stillinger}(1963)}]{Stillinger_rigorous_1963}%
  \BibitemOpen
  \bibfield  {author} {\bibinfo {author} {\bibfnamefont {F.~H.}\ \bibnamefont
  {Stillinger}, \bibfnamefont {Jr.}},\ }\href {\doibase 10.1063/1.1776907}
  {\bibfield  {journal} {\bibinfo  {journal} {J. Chem. Phys.}\ }\textbf
  {\bibinfo {volume} {38}},\ \bibinfo {pages} {1486} (\bibinfo {year}
  {1963})}\BibitemShut {NoStop}%
\bibitem [{\citenamefont {{ten Wolde}}\ \emph {et~al.}(1999)\citenamefont {{ten
  Wolde}}, \citenamefont {{Ruiz-Montero}},\ and\ \citenamefont
  {Frenkel}}]{tenWolde_numerical_1999}%
  \BibitemOpen
  \bibfield  {author} {\bibinfo {author} {\bibfnamefont {P.~R.}\ \bibnamefont
  {{ten Wolde}}}, \bibinfo {author} {\bibfnamefont {M.~J.}\ \bibnamefont
  {{Ruiz-Montero}}}, \ and\ \bibinfo {author} {\bibfnamefont {D.}~\bibnamefont
  {Frenkel}},\ }\href {\doibase 10.1063/1.477799} {\bibfield  {journal}
  {\bibinfo  {journal} {J. Chem. Phys.}\ }\textbf {\bibinfo {volume} {110}},\
  \bibinfo {pages} {1591} (\bibinfo {year} {1999})}\BibitemShut {NoStop}%
\bibitem [{\citenamefont {Wang}\ and\ \citenamefont
  {Frenkel}(2005)}]{Wang_pore_2005}%
  \BibitemOpen
  \bibfield  {author} {\bibinfo {author} {\bibfnamefont {Z.-J.}\ \bibnamefont
  {Wang}}\ and\ \bibinfo {author} {\bibfnamefont {D.}~\bibnamefont {Frenkel}},\
  }\href {\doibase 10.1063/1.2060666} {\bibfield  {journal} {\bibinfo
  {journal} {J. Chem. Phys.}\ }\textbf {\bibinfo {volume} {123}},\ \bibinfo
  {pages} {154701} (\bibinfo {year} {2005})}\BibitemShut {NoStop}%
\bibitem [{\citenamefont {Caupin}(2005)}]{Caupin_liquidvapor_2005}%
  \BibitemOpen
  \bibfield  {author} {\bibinfo {author} {\bibfnamefont {F.}~\bibnamefont
  {Caupin}},\ }\href {\doibase 10.1103/PhysRevE.71.051605} {\bibfield
  {journal} {\bibinfo  {journal} {Physical Review E}\ }\textbf {\bibinfo
  {volume} {71}},\ \bibinfo {pages} {051605} (\bibinfo {year}
  {2005})}\BibitemShut {NoStop}%
\bibitem [{\citenamefont {Cole}\ and\ \citenamefont
  {Saam}(1974)}]{Cole_excitation_1974}%
  \BibitemOpen
  \bibfield  {author} {\bibinfo {author} {\bibfnamefont {M.~W.}\ \bibnamefont
  {Cole}}\ and\ \bibinfo {author} {\bibfnamefont {W.~F.}\ \bibnamefont
  {Saam}},\ }\href {\doibase 10.1103/PhysRevLett.32.985} {\bibfield  {journal}
  {\bibinfo  {journal} {Phys. Rev. Lett.}\ }\textbf {\bibinfo {volume} {32}},\
  \bibinfo {pages} {985} (\bibinfo {year} {1974})}\BibitemShut {NoStop}%
\bibitem [{\citenamefont {Saam}\ and\ \citenamefont
  {Cole}(1975)}]{Saam_excitations_1975}%
  \BibitemOpen
  \bibfield  {author} {\bibinfo {author} {\bibfnamefont {W.~F.}\ \bibnamefont
  {Saam}}\ and\ \bibinfo {author} {\bibfnamefont {M.~W.}\ \bibnamefont
  {Cole}},\ }\href {\doibase 10.1103/PhysRevB.11.1086} {\bibfield  {journal}
  {\bibinfo  {journal} {Phys. Rev. B}\ }\textbf {\bibinfo {volume} {11}},\
  \bibinfo {pages} {1086} (\bibinfo {year} {1975})}\BibitemShut {NoStop}%
\bibitem [{\citenamefont {Kesselring}\ \emph {et~al.}(2012)\citenamefont
  {Kesselring}, \citenamefont {Franzese}, \citenamefont {Buldyrev},
  \citenamefont {Herrmann},\ and\ \citenamefont
  {Stanley}}]{Kesselring_nanoscale_2012}%
  \BibitemOpen
  \bibfield  {author} {\bibinfo {author} {\bibfnamefont {T.~A.}\ \bibnamefont
  {Kesselring}}, \bibinfo {author} {\bibfnamefont {G.}~\bibnamefont
  {Franzese}}, \bibinfo {author} {\bibfnamefont {S.~V.}\ \bibnamefont
  {Buldyrev}}, \bibinfo {author} {\bibfnamefont {H.~J.}\ \bibnamefont
  {Herrmann}}, \ and\ \bibinfo {author} {\bibfnamefont {H.~E.}\ \bibnamefont
  {Stanley}},\ }\href {\doibase 10.1038/srep00474} {\bibfield  {journal}
  {\bibinfo  {journal} {Scientific Reports}\ }\textbf {\bibinfo {volume} {2}}
  (\bibinfo {year} {2012}),\ 10.1038/srep00474}\BibitemShut {NoStop}%
\bibitem [{\citenamefont {Bruot}\ and\ \citenamefont
  {Caupin}(2016)}]{Bruot_curvature_2016}%
  \BibitemOpen
  \bibfield  {author} {\bibinfo {author} {\bibfnamefont {N.}~\bibnamefont
  {Bruot}}\ and\ \bibinfo {author} {\bibfnamefont {F.}~\bibnamefont {Caupin}},\
  }\href {\doibase 10.1103/PhysRevLett.116.056102} {\bibfield  {journal}
  {\bibinfo  {journal} {Phys. Rev. Lett.}\ }\textbf {\bibinfo {volume} {116}},\
  \bibinfo {pages} {056102} (\bibinfo {year} {2016})}\BibitemShut {NoStop}%
\bibitem [{\citenamefont {Puibasset}(2025)}]{Puibasset_molecularsized_2025}%
  \BibitemOpen
  \bibfield  {author} {\bibinfo {author} {\bibfnamefont {J.}~\bibnamefont
  {Puibasset}},\ }\href {\doibase 10.1063/5.0272545} {\bibfield  {journal}
  {\bibinfo  {journal} {The Journal of Chemical Physics}\ }\textbf {\bibinfo
  {volume} {163}},\ \bibinfo {pages} {034110} (\bibinfo {year}
  {2025})}\BibitemShut {NoStop}%
\end{thebibliography}%

\end{document}